 \pdfoutput=1
\documentclass[review
,onecolumn]{svjour3}       
\smartqed  

\usepackage{hyperref}
\usepackage{url}
\usepackage{graphics}
\usepackage[pdftex]{graphicx}
\usepackage{algorithmic}
\usepackage{algorithm}
\usepackage{fullpage}

\DeclareGraphicsExtensions{.jpg,.pdf,.tif,.png,.tiff,.eps}

\RequirePackage{fix-cm}

\usepackage[numbers,sort&compress]{natbib}
\usepackage{lineno}
\usepackage{algorithmic}
\usepackage[cmex10]{amsmath}
\usepackage{amssymb}
\usepackage{mdwmath}
\usepackage{mdwtab}
\usepackage{dblfloatfix}
\usepackage{setspace}

\usepackage{baskervald}
\usepackage{mathtools}

\usepackage{cancel}
\usepackage{caption}
\usepackage{subcaption}

\newcommand{\phiobs}{\varphi_{\textrm{obs}}}

\newcommand{\x}{\mathbf{ x}}

\newcommand{\z}{\mathbf{ z}}

\renewcommand{\omega}{{\boldsymbol\omega}}

\newcommand{\xp}{\mathbf{ x}^\prime}
\newcommand{\xpp}{\mathbf{ x}^{\prime\prime}}

\newcommand{\kk}{\mathbf{ k}}

\newcommand{\nablap}{\nabla_{\x^\prime}}
\newcommand{\Deltap}{\Delta_{\x^\prime}}
\renewcommand\d[1]{\:\textrm{d}#1}

\newcommand{\y}{\mathbf{ y}}

\renewcommand{\j}{\mathbf{ j}}

\graphicspath{{./figures/}}

\usepackage{appendix}

\begin{document}
\title{A path-integral approach to Bayesian inference for inverse problems using the semiclassical approximation}
\titlerunning{Field theoretic approach to Bayesian inference}
\journalname{Journal of Statistical Physics}
\date{\today}
\author{Joshua C. Chang \and Van M. Savage \and Tom Chou}
\institute{J.C. Chang \at
  Mathematical Biosciences Institute,The Ohio State University\\
Jennings Hall, 3rd Floor, 1735 Neil Avenue, Columbus, Ohio 43210\\
  \email{chang.1166@mbi.osu.edu} \\
  \and
  T. Chou \at
  UCLA Departments of Biomathematics and Mathematics\\
BOX 951766, Room 5309 Life Sciences, Los Angeles, CA 90095-1766 \\
  \email{tomchou@ucla.edu}\\
  \and
   V.M. Savage \at
   Santa Fe Institute, Santa Fe, NM 87501 \\
   UCLA Department of Biomathematics and Department of Ecology and Evolutionary Biology \\
   BOX 951766, Room 5205 Life Sciences, Los Angeles, CA 90095-1766 \\
  \email{vsavage@ucla.edu}
 }

\maketitle

\begin{abstract}

We demonstrate how path integrals often used in problems of
theoretical physics can be adapted to provide a machinery for
performing Bayesian inference in function spaces. Such inference comes
about naturally in the study of inverse problems of recovering continuous (infinite dimensional)
coefficient functions from ordinary or partial differential equations
(ODE, PDE), a problem which is typically ill-posed. Regularization of these
problems using $L^2$ function spaces (Tikhonov regularization) is
equivalent to Bayesian probabilistic inference, using a Gaussian
prior. The Bayesian interpretation of inverse problem regularization
is useful since it allows one to quantify and characterize error and degree
 of precision in the solution of inverse problems, as well as examine assumptions made in solving
the problem -- namely whether the subjective choice of regularization is compatible with prior knowledge.
Using path-integral formalism, Bayesian inference can be explored through various perturbative techniques,
such as the semiclassical approximation, which we use in this manuscript.
Perturbative path-integral approaches, while offering alternatives to computational approaches like
Markov-Chain-Monte-Carlo (MCMC), also provide natural starting points for
MCMC methods that can be used to refine approximations.
 In this manuscript, we  illustrate a path-integral formulation for inverse problems and
demonstrate it on an inverse problem in membrane biophysics as well as
 inverse problems in potential theories involving the
Poisson equation.

\end{abstract}


\section{Introduction}

One of the main conceptual challenges in solving inverse problems
results from the fact that most interesting inverse problems are not
well-posed. One often chooses a solution that is ``useful," or that
optimizes some regularity criteria. Such a task is commonly known
as \emph{regularization,} of which there are many variants.  One of
the most commonly used methods is \emph{Tikhonov Regularization}, or
$L^2$-penalized
regularization~\cite{engl2009inverse,engl1999convergence,neubauer1999tikhonov,hohage2008nonlinear,tikhonov1943stability}. 

Here we first demonstrate the concept behind Tikhonov regularization
using one of the simplest inverse problems, the interpolation
problem. Tikhonov regularization, when applied to interpolation,
solves the inverse problem of constructing a continuous function
$\varphi:\mathbb{R}^d\to\mathbb{R}$ from point-wise measurements
$\phiobs$ at positions $\{\x_m\}$ by seeking minima with respect to a
cost functional of the form

\begin{equation}
H[\varphi]
=\underbrace{\frac{1}{2}\sum_{m=1}^M \frac{1}{s_m^2}\left( \varphi(\x_m)-\phiobs(\x_m)\right)^2
}_{H_{\textrm{obs}}[\varphi]}+ \underbrace{\frac{1}{2}\sum_\alpha\gamma_\alpha\int
|D^\alpha \varphi|^2 \d\x}_{H_\textrm{reg}[\varphi]},
\end{equation}
where the constants $1/s_m^2,\gamma_\alpha>0$ are weighting
parameters, and $D^\alpha=\prod_{j=1}^d(-i\partial_{x_j})^{\alpha_j}$
is a differential operator of order
$\alpha=(\alpha_1,\ldots,\alpha_d)$.

Assuming $D^\alpha$ is isotropic and integer-ordered, it is possible
to invoke integration-by-parts to write $H[\varphi]$ in the quadratic
form
\begin{equation}
H[\varphi] = \frac{1}{2}\sum_{m=1}^M \frac{1}{s_m^2}\left( \varphi(\x_m)-\phiobs(\x_m)\right)^2 + \frac{1}{2}\int\varphi(\x) P(-\Delta)\varphi(\x) \d\x,
\label{eq:genericH}
\end{equation}
where $P(\cdot)$ is a polynomial of possibly infinite order, $\Delta$
is the Laplacian operator, and we have assumed that boundary terms
vanish. In the remainder of this work, we will focus on energy
functionals of this form. This expression is known in previous
literature as the \emph{Information Hamiltonian}~\cite{ensslin2009information}.

Using this form of regularization serves two primary purposes. First,
it selects smooth solutions to the inverse problem, with the amount of
smoothness controlled by $H_\textrm{reg}$.
For example, if only $H_\textrm{obs}$ is used, the
solution can be any function that connects the observations $\phiobs$
at the measured points $\x_j$, such as a piecewise affine
solution. Yet, such solutions may be physically unreasonable (not
smooth).
 Second, it transforms the
original inverse problem into a convex optimization problem that
possesses an unique
solution~\cite{bertero1980stability,engl1989convergence}. If all of
the coefficients of $P$ are non-negative, then the
pseudo-differential-operator $P(-\Delta)$ is
positive-definite~\cite{hormander2007analysis}, guaranteeing
uniqueness.  These features of Tikhonov regularization make it attractive;
however, one needs to make certain choices. In practical settings, one
will need to chose both the degree of the differential operator and
value of the parameters $\gamma_{\alpha}$. These two choices adjust the
trade-off between data agreement and regularity.

\subsection{Bayesian inverse problems}

The problem of parameter selection for regularization is
well-addressed in the context of \emph{Bayesian} inference, where
regularization parameters can be viewed probabilistically as
prior-knowledge of the solution. Bayesian inference over continuous function spaces has been applied to inverse problems in several contexts.
One of the first applications of Bayesian inference to inverse problems was in the study of
quantum inverse problems~\cite{lemm1999bayesian}, where it was noted that Gaussian priors could
be used to formulate field theories. Subsequently, variants of this
methodology have been used for model reduction~\cite{lieberman2010parameter} and applied to many interpolation problems and inverse problems in
fluid mechanics~\cite{cotter2009bayesian,stuart2010inverse,hoang2013determining}, geology~\cite{farmer2007bayesian,potsepaev2010application,martin2012stochastic},  cosmology~\cite{ensslin2009information,oppermann2011reconstructing}, and biology~\cite{heuett2012bayesian}.

There is a wealth of literature concerning the computational aspects of Bayesian inverse problems.
Many of these works on inverse problems are viewed through the framework and language of data assimilation through Markov Chain Monte Carlo approaches~\cite{quinn2011data,quinn2010state,bui2013computational,bui2013computational,petra2013computational}. 
Approximation methods based on sparsity have also been developed~\cite{schwab2012sparse}. Finally, there
is a large body of work on the theoretical aspects of maximum aposteriori inference for Bayesian
inverse problems including questions of existence of solutions and convergence to solutions~\cite{dashti2013map,lasanen2012non1,lasanen2012non2,stuart2010inverse,lasanen2007measurements}

\section{Field-theoretic formulation}

 Bayesian inference on $\varphi$
entails the construction of a probability density $\pi$ known as
the \emph{posterior distribution} $\pi(\varphi)$ which
obeys \emph{Bayes' rule,}

\begin{equation}
\pi(\varphi) = \frac{\overbrace{\Pr(\phiobs|\varphi)}^{\textrm{likelihood}}\overbrace{\Pr(\varphi)}^{\textrm{prior}}}{Z[0]}
\label{eq:posterior}
\end{equation}
where $Z[0]$ is the partition function or normalization factor. The posterior
density $\pi$ is a density in a space of functions.
The inverse problem is then investigated by computing the statistics
of the posterior probability density $\pi(\phi)$ through the evaluation of $Z[0]$.
 The solution of the inverse problem corresponds to the specific $\varphi$
 that maximizes $\pi(\phi)$, subject to prior knowledge 
 encoded in the prior probability density $\Pr(\varphi)$. This solution is known as the
  \emph{mean field} solution. The variance, or error,
of the mean field solution is found by computing the variance of the posterior distribution
about the mean field solution.

 This view of inverse problems also leads
naturally to the use of functional integration and perturbation
methods common in theoretical
physics~\cite{zee2005quantum,kardar2007statistical}. Use of the
probabilistic viewpoint allows for exploration of inverse problems
beyond mean field, with the chief advantage of providing a method for
uncertainty quantification. 

As shown in~\cite{lemm1999bayesian,farmer2007bayesian}, Tikhonov regularization has the
probabilistic interpretation of Bayesian inference with a Gaussian
prior distribution. That is, the regularization term in
Eq~\ref{eq:genericH} combines with the data term to specify a
posterior distribution of the form 

\begin{align}
\lefteqn{\pi(\varphi|\phiobs) = \frac{1}{Z[0]} e^{-H[\varphi]} }\nonumber\\
&= \frac{1}{Z[0]}\underbrace{\exp\left\{ - \sum_{m=1}^M \frac{1}{s_m^2}\left( \varphi(\x_m)-\phiobs(\x_m)\right)^2\right\}}_{\textrm{likelihood}\ (\exp\{-H_\textrm{obs}\})}\underbrace{\exp\left\{ -\frac{1}{2}\int \varphi(\x) P(-\Delta)\varphi(\x) \d\x \right\}}_{\textrm{prior}\ (\exp\{-H_\textrm{reg}\})}
\label{eq:pi}
\end{align}
where the partition function

\begin{equation}\label{eq:abstractfunctionalintegral}
Z[0]=\int \mathcal{D}\varphi e^{-H[\varphi]}
= \int\underbrace{\mathcal{D}\varphi
e^{-H_\textrm{reg}[\varphi]}}_{\textrm{d}W[\varphi]}e^{-H_\textrm{obs}[\varphi]}
\end{equation}
is a sum over the contributions of all functions in the separable
Hilbert space $\{ \varphi: H_\textrm{reg}[\varphi]<\infty\}$. This sum
is expressed as a \emph{path integral}, which is an integral over a
function space. The formalism for this type of integral came about
first from high-energy theoretical physics~\cite{feynman2012quantum}, and then found application
in nearly all areas of physics as well as in the representation of
both Markovian~\cite{chow2010path,graham1977path,peliti1985path}, and
non-Markovian~\cite{pesquera1983path,hanggi1989path} stochastic
processes. In the case of Eq.~\ref{eq:abstractfunctionalintegral},
where the field theory is real-valued and the operator $P(-\Delta)$ is
self-adjoint, a type of functional integral based on abstract Wiener
measure may be used~\cite{ito1961wiener}. The abstract Wiener measure $\textrm{d}W[\varphi]$ used
for Eq.~\ref{eq:abstractfunctionalintegral} subsumes the prior term
$H_\textrm{reg}$, and it is helpful to think of it as a Gaussian
measure over lattice points taken to the continuum limit. 

When the functional integral of the exponentiated energy functional
can be written in the form 

\begin{equation}\label{eq:gaussianform}
Z[0]=\int \mathcal{D}\varphi\exp\left\{-\frac{1}{2}\iint \varphi(\x)A(\x,\xp)\varphi(\xp)\d\x
\d\xp +\int b(\x)\varphi(\x)\d\x \right\},
\end{equation} 
then the probability density is Gaussian in function-space and the
functional integral of Eq.~\ref{eq:gaussianform} has the
solution~\cite{zee2005quantum} 

\begin{equation}\label{eq:integratedgaussian}
Z[0]= \exp\left\{\frac{1}{2} \iint b(\x)A^{-1}(\x,\xp)b(\xp)\d\x \d\xp
-\frac{1}{2}\log\det A\right\}.  
\end{equation} 
The operators $A(\x,\xp)$ and $A^{-1}(\x,\xp)$ are related through the
relationship
\begin{equation} \int A(\x,\xp)A^{-1}(\xp,\xpp)\d\xp=\delta(\x-\xpp).
\label{eq:inverserelationship}
\end{equation}

Upon neglecting $H_\textrm{obs}$, the functional integral of
Eq.~\ref{eq:abstractfunctionalintegral} can be expressed in the form
of Eq.~\ref{eq:gaussianform} with
$A(\x,\xp)=P(-\Delta)\delta(\x-\xp)$. The pseudo-differential-operator $P(-\Delta)$ acts as an
infinite-dimensional version of the inverse of a covariance matrix. It
encodes the a-priori spatial correlation, implying that values of the
function $\varphi$ are spatially correlated according to a correlation
function (Green's function)
$A^{-1}(\x,\y)=G(\x,\y):\mathbb{R}^d\times\mathbb{R}^d\to\mathbb{R}$ through the
relationship implied by Eq.~\ref{eq:inverserelationship},
$P(-\Delta)G(\x,\y)=\delta(\x-\y)$ so that
$G(\x,\y)= \left(\frac{1}{2\pi}\right)^d\int_{\mathbb{R}^d}
e^{-i\kk\cdot (\y-\x)}\frac{1}{P(|\kk|^2)}\d\kk$ where $P(|\kk|^2)$ is
the symbol of the pseudo-differential-operator $P(-\Delta)$.  It is
evident that when performing Tikhonov regularization, one should chose
regularization that is reflective of prior knowledge of correlations,
whenever available.

\subsection{Mean field inverse problems}
We turn now to the more-general problem, where one seeks recovery of a
scalar function $\xi$ given measurements of a coupled scalar function
$\varphi$ over interior points $\x_i$, and the relationship between the
measured and desired functions is given by a partial differential
equation

\begin{equation}
F(\varphi(\x),\xi(\x))=0\qquad \x\in\Omega\setminus\partial\Omega.
\end{equation}
As before, we regularize $\xi$ using knowledge of its spatial
correlation, and write a posterior probability density
\begin{align*}
\lefteqn{\pi[\varphi,\xi|\phiobs] = }\\
&\quad\frac{\delta\left(F(\varphi,\xi) \right)}{Z[0]}\exp\left\{-\frac{1}{2}\int\sum_{m=1}^M \delta(\x-\x_m) \frac{\left(\varphi(\x)-\phiobs(\x) \right)^2}{s_m^2} \d\x -\frac{1}{2}\int \xi(\x)P(-\Delta)\xi(\x)\d\x \right\}\boldsymbol,
\end{align*}
where we have used the Dirac-delta function $\delta$ to specify that
our observations are taken with noise $s_m^2$ at certain positions
$\x_m$, and an infinite-dimensional delta functional
$\boldsymbol\delta$ to specify that $F(\varphi,\xi)=0$
everywhere. Using the inverse Fourier-transformation, one can
represent $\boldsymbol\delta$ in path-integral form as
$
\boldsymbol\delta\left(F(\varphi,\xi) \right)=\int \mathcal{D}\lambda e^{-i\int \lambda(x)F(\varphi(\x),\xi(\x))\d\x },
$
where $\lambda(\x)$, is a Fourier wavevector. The reason for this
notation will soon be clear. We now have a posterior probability
distribution of three functions $\varphi,\xi,\lambda$ of the form

\begin{equation}
\pi[\varphi,\xi,\lambda(\x)|\phiobs]=\frac{1}{Z[0]} \exp\left\{-H[\varphi,\xi,\lambda] \right\} ,
\end{equation}
where the \emph{partition functional} is

\begin{equation}
Z[0] = \iiint \mathcal{D}\varphi\mathcal{D}\xi\mathcal{D}\lambda\exp\left\{-H[\varphi,\xi,\lambda] \right\},
\end{equation}
and the \emph{Hamiltonian} 
\begin{align}
 \lefteqn{H[\varphi,\xi,\lambda;\phiobs]=\quad\frac{1}{2}\int\sum_{m=1}^M \delta(\x-\x_m)\frac{
 (\varphi(\x)-\phiobs(\x))^2}{s_m^2}\d\x }\nonumber\\
 &\quad+\frac{1}{2}\int \xi(\x)P(-\Delta)\xi(\x)\d\x +
 i\int \lambda(\x)F(\varphi,\xi) \d\x,
\label{eq:constrainedenergy}
\end{align}
is a functional of $\varphi,\xi$, and the Fourier wave vector
$\lambda(\x)$. Similar Hamiltonians, providing a probabilistic model for 
data in the context of inverse problems, have appeared in previous literature
~\cite{ensslin2009information,lemm1999bayesian,stuart2010inverse}, 
where they have been referred to as \emph{Information Hamiltonians.}

    Maximization of the posterior probability distribution, also known
as Bayesian maximum a posteriori estimation (MAP) inference, is performed by
minimization of the corresponding energy functional
(Eq.~\ref{eq:constrainedenergy}) with respect to the functions
$\varphi,\xi,\lambda$. One may perform this inference by solving the associated
Euler-Lagrange equations

\begin{align}
P(-\Delta)\xi + \frac{\delta}{\delta \xi(\x)}\int {\lambda}(\x)F(\varphi,\xi) \d\x    &= 0,\\
\sum_{n=1}^M \delta(\x-\x_n)(\varphi(\x)-\phiobs(\x))+ \frac{\delta}{\delta \varphi(\x)}\int {\lambda}(\x)F(\varphi,\xi) \d\x    &= 0  \\
F(\varphi,\xi) &=0,
\end{align}
where $\lambda(\x)$ here serves the role of a Lagrange
multiplier. Solving this system of partial differential equations
simultaneously allows one to arrive at the solution to the original
Tikhonov-regularized inverse problem. Now, suppose one is interested
in estimating the precision of the given solution. The field-theoretic
formulation of inverse problems provides a way of doing so.

\subsection{Beyond mean-field -- semiclassical approximation}

The functions $\varphi,\xi,\lambda:\mathbb{R}^d\to\mathbb{R}$ each
constitute scalar fields\footnote{We will use Greek letters to denote
fields}. \emph{Field theory} is the study of statistical properties of
such fields through evaluation of an associated \emph{path integral} (functional integral). 
 Field theory applied to Bayesian inference has appeared in prior literature
 under the names Bayesian Field theory~\cite{lemm1999bayesian,farmer2007bayesian,stuart2010inverse},
  and Information Field Theory~\cite{ensslin2009information}. 
  
 In general, field theory deals with functional integrals of the form

\begin{equation}
Z[J]=\int \mathcal{D}\varphi \exp\Bigg\{-\underbrace{\left[\frac{1}{2}\iint \varphi(\x) A(\x,\xp)\varphi(\xp)\d\x \d\xp + \int V[\varphi(\x)]\d\x \right]}_{H[\varphi]}+ \int J(\x)\varphi(\x) \d{\x}  \Bigg\},\label{eq:generatingfunctional}
\end{equation}
where the Hamiltonian of interest is recovered when the \emph{source}
$J=0$, and the potential function $V$ is nonlinear in
$\varphi$. Assuming that after non-dimensionalization, $V[\varphi]$ is
relatively small in comparison to the other terms, one is then able to
expand the last term in formal Taylor series so that after completing
the Gaussian part of the integral as in
Eq.~\ref{eq:integratedgaussian},

\begin{align}
\lefteqn{Z[J] = \int\mathcal{D}\varphi \Bigg\{ \underbrace{\exp\left[-\frac{1}{2}\iint \varphi(\x) A(\x,\xp)\varphi(\xp)\d\x \d\xp + \int J(\x)\varphi(\x) \d{\x}\right]}_\textrm{Gaussian}} \nonumber\\
&\qquad\qquad\times\left(1-\int V[\varphi]\d\x+\ldots \right) \Bigg\}\nonumber\\
&\propto\exp\left[-{V\left(\frac{\delta}{\delta J} \right)}\right]\exp\left({\frac{1}{2}\iint J(\x)A^{-1}(\x,\xp)J(\xp)\d\x \d\xp}\right).
\end{align}

In this way, $Z[J]$ can be expressed in series form as moments of a
Gaussian distribution. The integral is of interest because one can use
it to recover moments of the desired field through functional
differentiation,

\begin{equation}
\left\langle \prod_k \varphi(\x_k) \right\rangle = \left.\frac{1}{Z[0]}\prod_k 
\frac{\delta}{\delta J(\x_k)} Z[J]\right|_{J=0}.
\label{eq:moments}
\end{equation}
This approach is known as the \emph{weak-coupling approach}~\cite{zee2005quantum}.
For this expansion to hold, however, the external potential $V$ must
be small in size compared to the quadratic term. This assumption is
not generally valid during Tikhonov regularization, as common rules of
thumb dictate that the data fidelity and the regularization term
should be of similar order of
magnitude~\cite{anzengruber2010morozov,scherzer1993use}.  Another
perturbative approach -- the one that we will take in this manuscript
-- is to expand the Hamiltonian in a functional Taylor series
\begin{equation}
H[\varphi] = H[\varphi^\star] +  \frac{1}{2}\iint 
\frac{\delta^2 H[\varphi^\star]}{\delta \varphi(\x)\varphi(\xp)}
(\varphi(\x)-\varphi^\star(\x))(\varphi(\xp)-\varphi^\star(\xp))\d\x
\d\xp+\ldots\label{eq:happrox}
\end{equation}
about its extremal point $\varphi^\star$. 
To the second order (as shown), the expansion is known as the \emph{semiclassical approximation}~\cite{heller1981frozen} which
 provides an approximate Gaussian density for the field $\varphi$. Corrections to the semiclassical
 expansion can be evaluated by continuing this expansion to higher orders, where
 evaluation of the functional integral can be aided by the use of Feynman diagrams~\cite{feynman2012quantum}. 

\subsection{Monte-Carlo for refinement of approximations}\label{sec:montecarlo}

 The Gaussian approximation is useful because Gaussian densities are easy to sample. 
 One may sample a random field $\varphi(\x)$
from a Gaussian distribution with inverse-covariance $A(\x,\xp)$ by solving the stochastic differential
equation
\begin{equation}
\frac{1}{2}\int A(\x,\xp)\varphi(\xp)\d\xp = \eta(\x),\label{eq:Taylor}
\end{equation}
where $\eta$ is the unit white noise process which has mean $
\left\langle\eta(\x)\right\rangle = 0,
$
and spatial correlation
$
\left\langle\eta(\x)\eta(\xp)\right\rangle = \delta(\x-\xp).
$ With the ability to sample from the approximating Gaussian distribution
 of Eq.~\ref{eq:happrox}, one may use Monte-Carlo simulation to sample from the
true distribution by weighting the samples obtained from the Gaussian
distribution. Such an approach is known as \emph{importance sampling}~\cite{liu2008monte},
where samples $\varphi_i$ are given importance weights $w_i$ according
to the ratio $ w_i= \exp\left( -H_\textrm{approx} +
H_\textrm{true} \right)/\sum_j w_j.  $ Statistics of $\varphi$ may
then be calculated using the weighted samples; for instance
expectations can be approximated as $
\left\langle g(\varphi(\x))\right\rangle \approx \sum_i w_i g(\varphi_i(\x)).
$
Using this method, one can refine the original estimates of the statistics of $\varphi$.

\section{Examples}   

\subsection{Interpolation of the height of a rigid membrane or plate}\label{sec:interp}

We first demonstrate the field theory for inverse problems on an
interpolation problem where one is able to determine the regularizing
differential operator based on prior knowledge. This example
corresponds to the interpolation example mentioned in the
Introduction. Consider the problem where one is attempting to identify
in three-dimensions the position of a membrane. For simplicity, we
assume that one is interested in obtaining the position of the
membrane only over a restricted spatial domain, where one can use the
Monge parameterization to reduce the problem to two-dimensions and
define the height of the membrane $\varphi:\mathbb{R}^2\to\mathbb{R}$.

Suppose one is able to measure the membrane in certain spatial
locations $\{\x_m\}$, but one seeks to also interpolate the membrane in regions
that are not observable. Physically, models for fluctuations in
membranes are well known, for instance the Helfrich
free-energy~\cite{evans2003interactions} suggests that one should use
a regularizing differential operator

\begin{equation}\label{eq:helfrichdo}
P(-\Delta)=\beta(\kappa\Delta^2-\sigma\Delta)\qquad \beta,\sigma,\kappa>0,
\end{equation}
where $\sigma$ and $\kappa$ are the membrane tension and bending
rigidity, respectively.  The Hamiltonian associated with the Helfrich
operator is

\begin{equation}
H[\varphi; \phiobs]
= \frac{1}{2}\int \sum_{m=1}^M \frac{\delta(\x-\x_m)}{s_m^2}(\varphi(\x)-\phiobs(\x))^2\d\x
+ \frac{1}{2}\int \varphi(\x)P(-\Delta)\varphi(\x)\d\x, 
\end{equation}
and the mean-field solution for $\varphi$ corresponds to the extremal
point of the Hamiltonian, which is the solution of the corresponding Euler-Lagrange
equation

\begin{equation}
\frac{\delta H}{\delta \varphi} = \sum_{m=1}^M \frac{\delta(\x-\x_m)}{s_m^2}(\varphi(\x)-\phiobs(\x))+P(-\Delta)\varphi(\x) = 0.
\end{equation}
To go beyond mean-field, one may compute statistics of the probability
distribution $\Pr(\varphi) \propto e^{-H[\varphi]}$, using the
generating functional which is expressed as a functional integral

{
\begin{align}
Z[J] &\propto  \int \mathcal{D}\varphi \exp\Bigg\{- \frac{1}{2}\iint\varphi(\x)\underbrace{\left[ \delta(\x-\xp) \sum_{m=1}^M \frac{\delta(\xp-\x_m)}{s_m^2}+P(-\Delta)\delta(\x-\xp) \right]}_{A(\x,\xp)}\varphi(\xp)\d\x \d\xp \nonumber\\
&\qquad\qquad\qquad+ \int \left[  \sum_{m=1}^M \frac{\phiobs(\x)\delta(\x-\x_m)}{s_m^2}+J(\x)\right]\varphi(\x)\d\x \Bigg\}, \label{eq:zjinterp}
\end{align}}
where we have completed the square.  According to
Eq.~\ref{eq:integratedgaussian}, Eq.~\ref{eq:zjinterp} has the
solution
\begin{equation}
Z[J] \propto\exp\left\{\frac{1}{2}\iint J(\x)A^{-1}(\x,\xp)J(\xp)\d\xp \d\x + \int J(\x)\sum_{m=1}^M \frac{\phiobs(\x_m)A^{-1}(\x,\x_m)}{s_m^2} \d\x  \right\}.
\label{eq:zjplate}
\end{equation}
Through functional differentiation of Eq.~\ref{eq:zjplate},
Eq.~\ref{eq:moments} implies that the mean-field solution is

\begin{equation}\label{eq:memmean}
\left\langle \varphi(\x) \right\rangle =  \sum_{m=1}^M\frac{\phiobs(\x_m) A^{-1}(\x,\x_m)}{s_m^2} ,
\end{equation}
and variance in the solution is

\begin{equation}\label{eq:memvar}
\Big\langle \varphi(\x)- \left\langle \varphi(\x) \right\rangle,\varphi(\xp)-\left\langle \varphi(\xp) \right\rangle\Big\rangle =A^{-1}(\x,\xp).
\end{equation}

To solve for these quantities, we compute the operator $A^{-1}$, which
according to Eq.~\ref{eq:inverserelationship}, satisfies the partial
differential equation

\begin{equation}
 \sum_{m=1}^M \frac{\delta(\x_m-\x)}{s^2_m}A^{-1}(\x,\x^{\prime\prime})+P(-\Delta) A^{-1}(\x,\x^{\prime\prime}) = \delta(\x-\x^{\prime\prime}).
\end{equation}
Using the Green's function for $P(-\Delta)$, 

\begin{equation}
G(\x,\xp)
= \frac{-1}{2\pi\beta\sigma}\left[ \log\left(|\x-\xp|\right)+K_0\left(\sqrt{\frac{\sigma}{\kappa}}|\x-\xp| \right)\right], 
\end{equation}
we find

\begin{align}
A^{-1}(\x,\x^{\prime\prime})
&= \overbrace{G(\x,\xpp)}^{\textrm{known}}
- \sum_{m=1}^M \frac{ \overbrace{G(\x,\x_m)}^{\textrm{known}}\overbrace{A^{-1}(\x_m,\xpp)}^{\textrm{unknown}}}{s_m^2}.
\end{align}

To calculate $A^{-1}(\x,\xp),$ we need $A^{-1}(\x_m,\xp)$, for
$m\in\{1,\ldots,M\}$. Solving for each of these simultaneously yields the equation

\begin{equation}\label{eq:memAinv}
A^{-1}(\x,\xp) = G(\x,\xp) - \mathbf{G}_s(\x)\left( \mathbf{I} +\boldsymbol{\Lambda}\right)^{-1}\mathbf{G}(\xp),
\end{equation}
where
$\mathbf{G}_s(\x)\equiv\left[ \frac{G(\x,\x_1)}{s_1^2}, \frac{G(\x,\x_2)}{s_2^2},\ldots,\frac{G(\x,\x_M)}{s_M^2}\right]$, $\mathbf{G}(\x)\equiv\left[ {G(\x,\x_1)}, {G(\x,\x_2)},\ldots,{G(\x,\x_M)}\right]$, and $\mathbf{\Lambda}_{ij}\equiv G(\x_i,\x_j)/s_i^2$.

\begin{figure}[h]\centering
\begin{subfigure}[b]{0.32\textwidth}
\includegraphics[width=\textwidth]{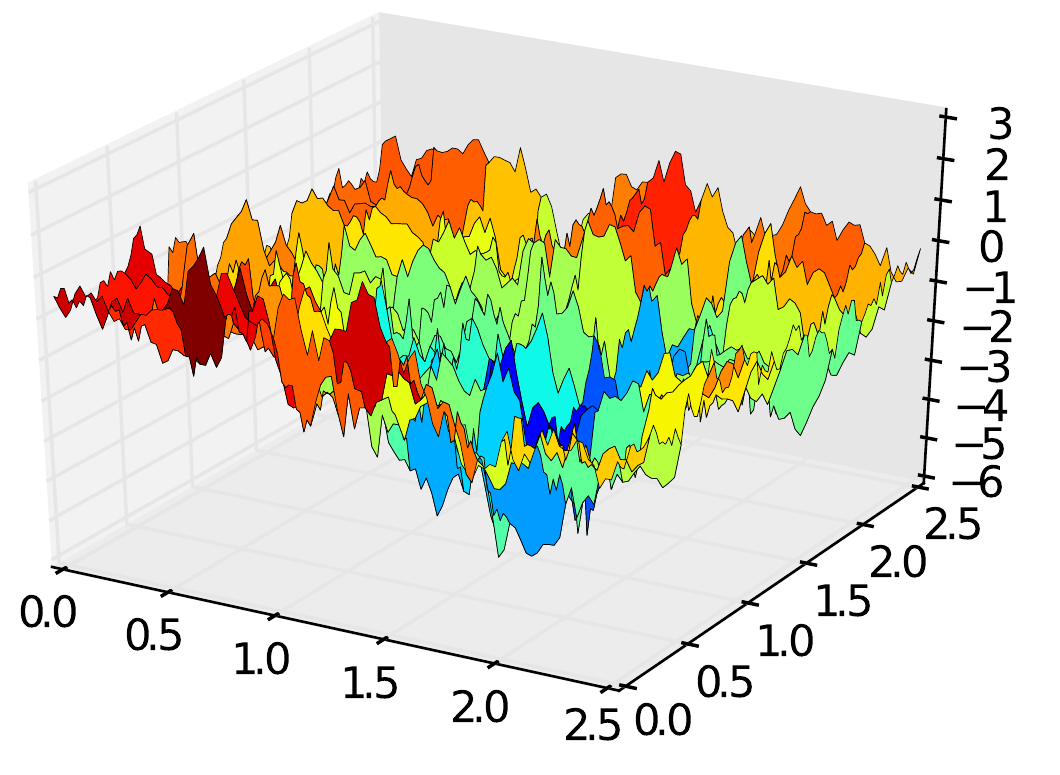}
\caption{$\varphi(\x)$}
\label{fig:membranetruth}
\end{subfigure}
\begin{subfigure}[b]{0.32\textwidth}
\includegraphics[width=\textwidth]{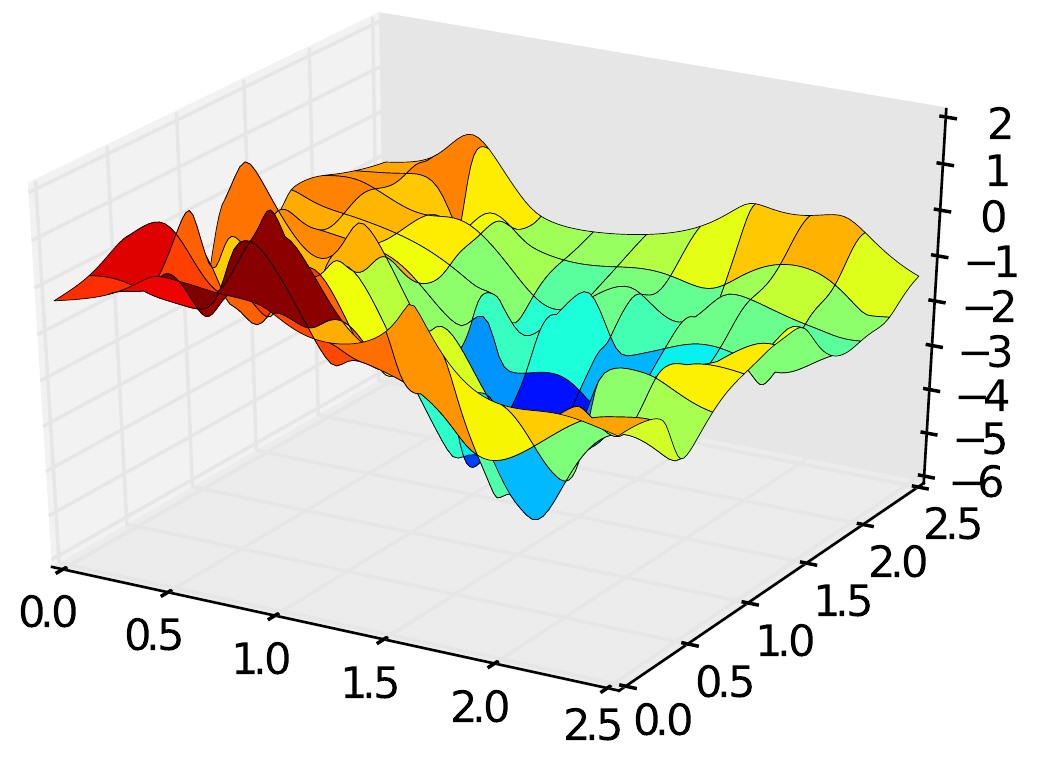}
\caption{$\langle\varphi(\x)\rangle$}
\label{fig:membrane}
\end{subfigure}
\begin{subfigure}[b]{0.32\textwidth}
\includegraphics[width=\textwidth]{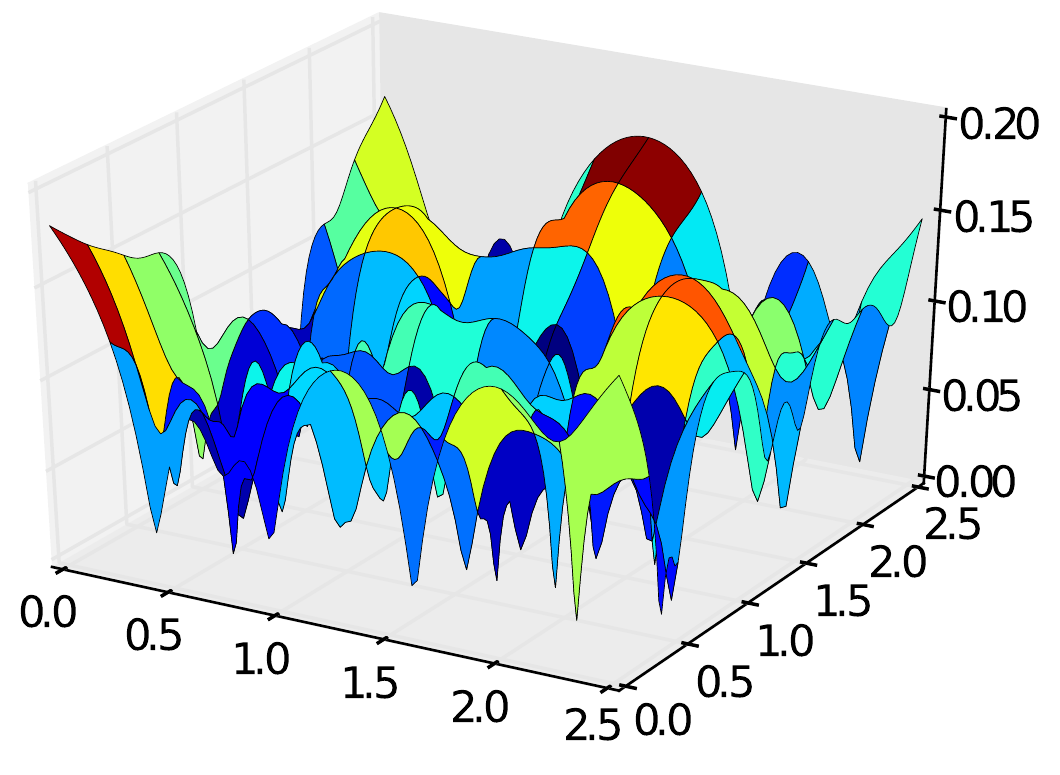}
\caption{$\sqrt{\langle(\varphi(\x) -\langle\varphi(\x)\rangle)^2\rangle}$}
\label{fig:membraneerr}
\end{subfigure}
\caption{\textbf{Interpolation of a membrane.} 
(\subref{fig:membranetruth}) A simulated membrane undergoing 
thermal fluctuations is the object of reconstruction. 
 (\subref{fig:membrane}) Mean-field reconstruction of the membrane using
$100$ randomly-placed measurements with
noise. (\subref{fig:membraneerr}) Pointwise standard error in the
reconstruction of the membrane. Parameters used: $\sigma=10^{-2}, \beta =
10^3, \kappa=10^{-4},s_m=10^{-2}$.  }
\label{fig:fig1}
\end{figure}

Fig.~\ref{fig:fig1} shows an example of the use of the Helfrich free
energy for interpolation. A sample of a membrane undergoing thermal
fluctuations was taken as the object of recovery. Uniformly, $100$
randomly-placed, noisy observations of the height of the membrane were
taken. The mean-field solution for the position of the membrane and
the standard error in the solution are presented. The standard error
is not uniform and dips to approximately the measurement error
at locations where measurements were taken.


\subsection{Source recovery for the Poisson equation}\label{sec:source}

Now consider an example where the function to be  recovered is not
directly measured. This type of inverse problem often arises when considering 
the Poisson equation in isotropic medium:

\begin{equation}
\Delta \varphi(\x) = \rho(\x).
\end{equation}
Measurements of $\varphi$ are taken at points
$\{\x_m\}$ and the objective is to recover the source function $\rho(\x)$.
Previous researchers have explored the use of Tikhonov regularization
to solve this problem~\cite{hon2010inverse,alves2008recovering}; here
we quantify the precision of such solutions.

Making the assumption that $\rho$ is correlated according to the
Green's function of the pseudo-differential-operator $P(-\Delta)$, we
write the Hamiltonian
\begin{align}
H[\varphi,\rho,\lambda; \phiobs]&=\frac{1}{2}\int \sum_{m=1}^M \frac{\delta(\x-\x_m)}{s_m^2}(\varphi(\x)-\phiobs(\x))^2\d\x +\frac{1}{2}\int \rho(\x) P(-\Delta) \rho(\x) \d\x \nonumber\\
&\qquad + i\int \lambda(\x)\left(\Delta \varphi(\x)-\rho(\x) \right)\d\x. \label{eq:poissonenergy0}
\end{align}
The extremum of $H[\varphi,\rho,\lambda; \phiobs]$ occurs at
$(\varphi^\star,\rho^\star)$, which are found through the
corresponding Euler-Lagrange equations $\left(\frac{\delta
H}{\delta \varphi}=0,\frac{\delta H}{\delta \rho} = 0,\frac{\delta
H}{i\delta \lambda}=0 \right)$,
\begin{align}
 \sum_{m=1}^M\frac{\delta(\x-\x_m)}{s_m^2}(\varphi^\star(\x)-\phiobs(\x))+P(-\Delta)\Delta^2 \varphi^\star(\x)&= 0, \nonumber \\
\rho^\star&= \Delta \varphi^\star .
\label{eq:sourceeuler}
\end{align}
In addition to the extremal solution, we can also evaluate how
precisely the source function has been recovered by considering the
probability distribution given by the exponentiated Hamiltonian,

\begin{align}
\lefteqn{\pi(\rho(\x) | \{\phiobs(\x_i)\}) =\frac{1}{Z[0]} } \nonumber\\
& \times\exp\left\{-\frac{1}{2}\int \sum_{m=1}^M \frac{\delta(\x-\x_m)}{s_m^2}(\varphi(\x)-\phiobs(\x))^2\d\x -\frac{1}{2}\int \Delta\varphi(\x) P(-\Delta)  \Delta\varphi(\x) \d\x  \right\},
\end{align}
where we have integrated out the $\lambda$ and $\rho$ variables by
making the substitution $\rho=\Delta \varphi$. To compute the
statistics of $\varphi$, we first compute $Z[J]$, the generating
functional which by Eq.~\ref{eq:integratedgaussian} has the solution

\begin{align}
 \lefteqn{Z[J]\propto \exp\Bigg\{\frac{1}{2}\iint\Delta J(\x)
 A^{-1}(\x,\x^\prime)\Deltap J(\x^\prime)\d\x^\prime \d\x} \nonumber\\
 &\qquad + \int
 J(\x)\Delta\sum_{m=1}^M \frac{\phiobs(\x_m)A^{-1}(\x,\x_m)}{s_m^2}\d\x \Bigg\},\label{eq:poissonZJ1}
\end{align}
where 

\begin{equation}
A(\x,\xp) = \Delta^2P(-\Delta)\delta(\x-\xp) + \delta(\x-\xp)\sum_{m=1}^M \frac{\delta(\x-\x_m)}{s_m^2} 
\end{equation}
and $A^{-1}$ is defined as in Eq.~\ref{eq:inverserelationship}.  The
first two moments have the explicit solution given by the generating
functional,

\begin{align*}
\left.\frac{\delta Z[J]}{\delta J(\x)}\right|_{J=0} & = \left( \sum_{m=1}^M\frac{\phiobs(\x_m)\Delta A^{-1}(\x,\x_m)}{s_m^2}  \right)Z[0]
\end{align*}

\begin{align*}
\lefteqn{\left.\frac{\delta^2 Z[J]}{\delta J(\x)\delta J(\xp)}\right|_{J=0} =Z[0]\Bigg[ \Delta\Deltap A^{-1}(\x,\xp)}\nonumber\\
& + \left( \sum_{m=1}^M\frac{\phiobs(\x_m)\Delta A^{-1}(\x,\x_m)}{s_m^2} \right) \left( \sum_{k=1}^M\frac{\phiobs(\x_k)\Deltap A^{-1}(\xp,\x_k)}{s_k^2} \right)\Bigg].
\end{align*}
These formulae imply that our mean-field source has the solution
\begin{equation}\label{eq:sourcemean}
\left\langle \rho(\x) \right\rangle =  \sum_{m=1}^M\frac{\phiobs(\x_m)\Delta A^{-1}(\x,\x_m)}{s_m^2} ,
\end{equation}
subject to the weighted unbiasedness condition $\sum_m \varphi(\x_m)/s_m^2 = \sum_m \phiobs(\x_m)/s_m^2$,
and the variance in the source has the solution
\begin{equation}\label{eq:sourcevariance}
\Big\langle \rho(\x)- \left\langle \rho(\x) \right\rangle,\rho(\xp)-\left\langle \rho(\xp) \right\rangle\Big\rangle =  \Delta\Deltap A^{-1}(\x,\xp).
\end{equation}
The inverse operator $A^{-1}$ is solved in the same way as in the
previous section, yielding for the fundamental solution $G$ satisfying
$P(-\Delta)\Delta^2 G(\x) = \delta(\x)$,

\begin{equation}
A^{-1}(\x,\xp) = G(\x,\xp) - \mathbf{G}_s(\x)\left( \mathbf{I} +\boldsymbol{\Lambda}\right)^{-1}\mathbf{G}(\xp),
\end{equation}
where $\mathbf{G},\mathbf{G}_s$ and $\boldsymbol\Lambda$ are defined
as they are in Eq.~\ref{eq:memAinv}.



\begin{figure}
\centering
\begin{subfigure}[b]{0.32\textwidth}
\subcaption{$\rho(\x)$}
\includegraphics[height=9pc]{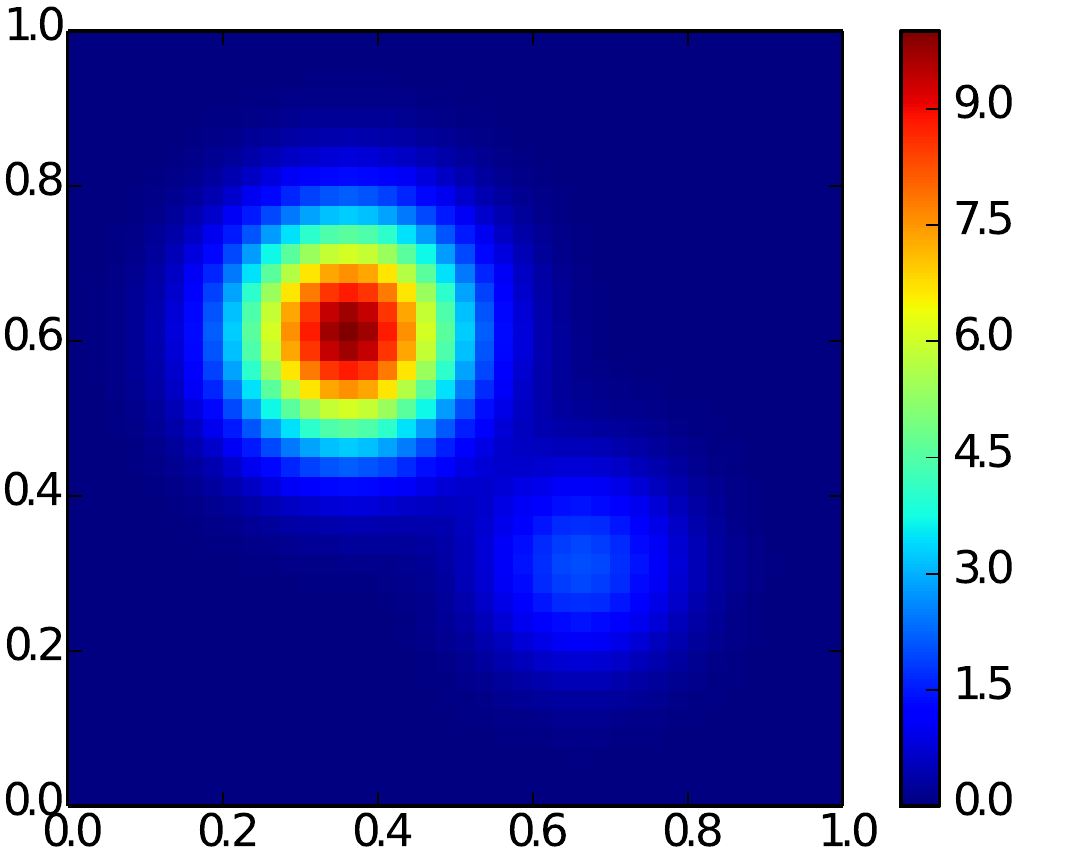}
\label{fig:fig2source}
\end{subfigure}
\begin{subfigure}[b]{0.32\textwidth}
\subcaption{$\varphi(\x)$}
\includegraphics[height=9pc]{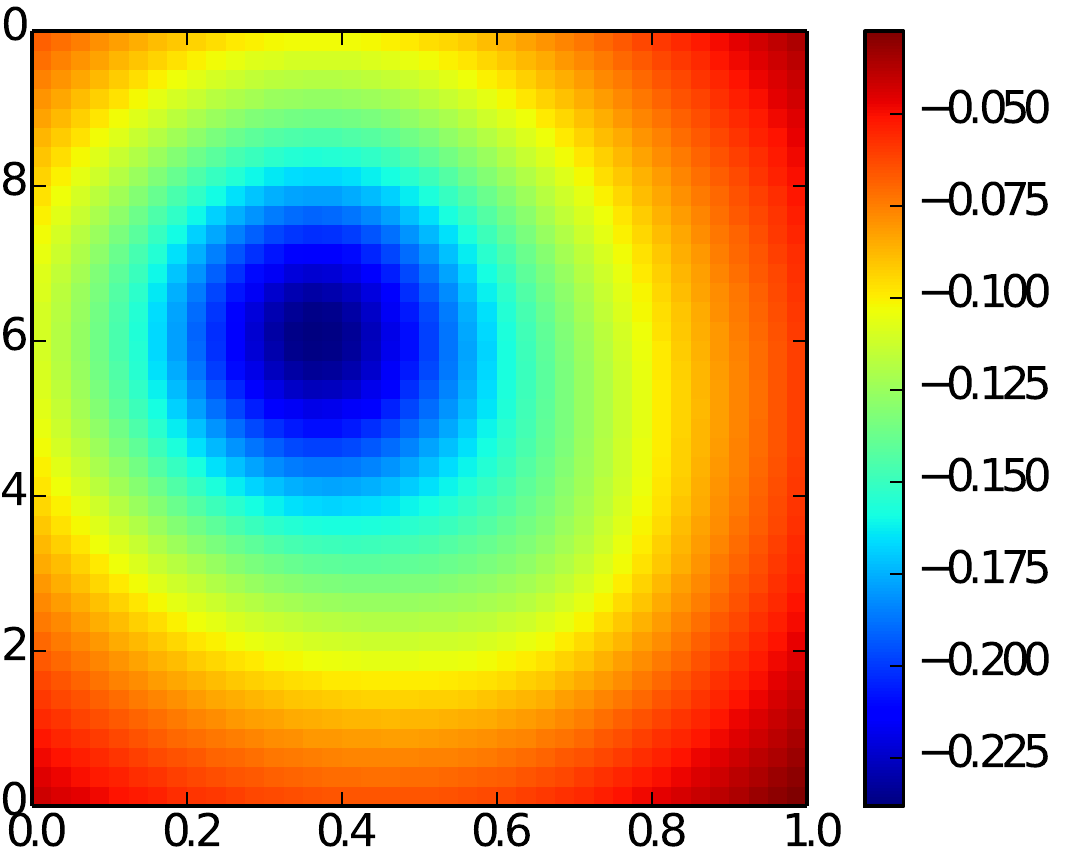}
\label{fig:fig2varphi}
\end{subfigure}
\begin{subfigure}[b]{0.32\textwidth}\centering
\subcaption{ $\phiobs(\x)$}
\includegraphics[height=9pc]{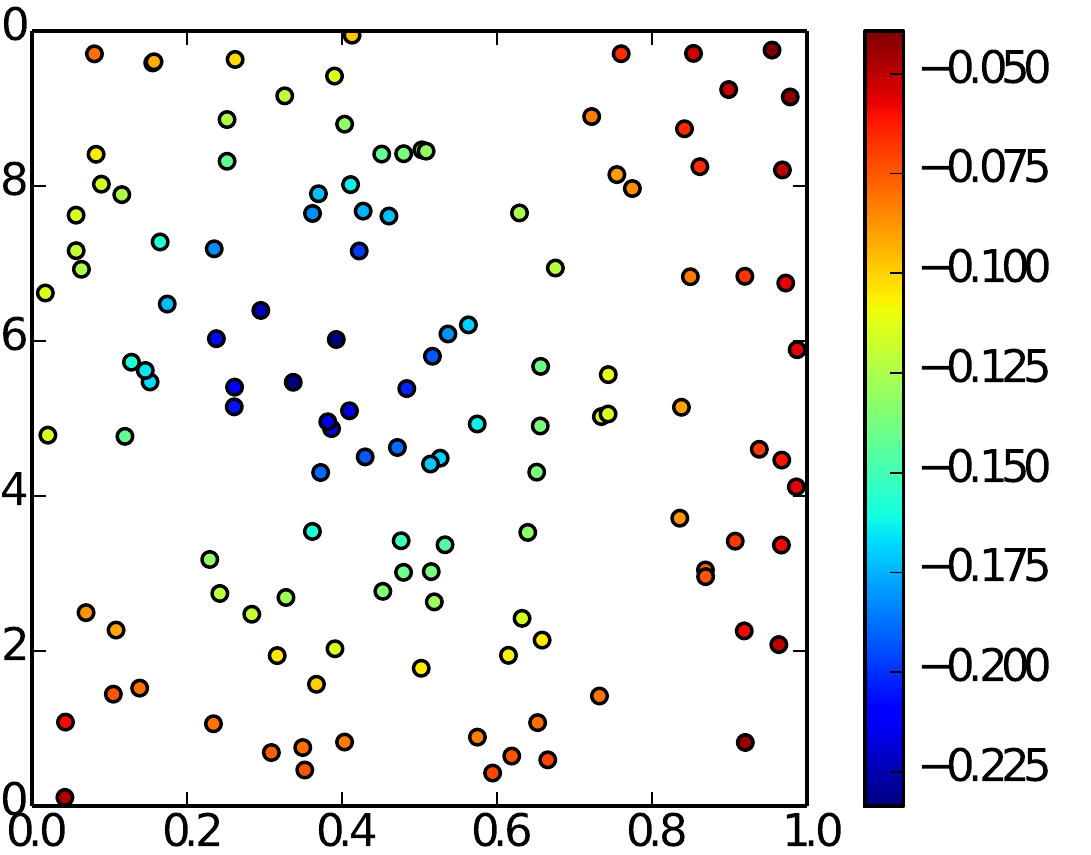}
\label{fig:fig2obs}
\end{subfigure}
%

\begin{subfigure}[b]{0.32\textwidth}
\includegraphics[height=9pc]{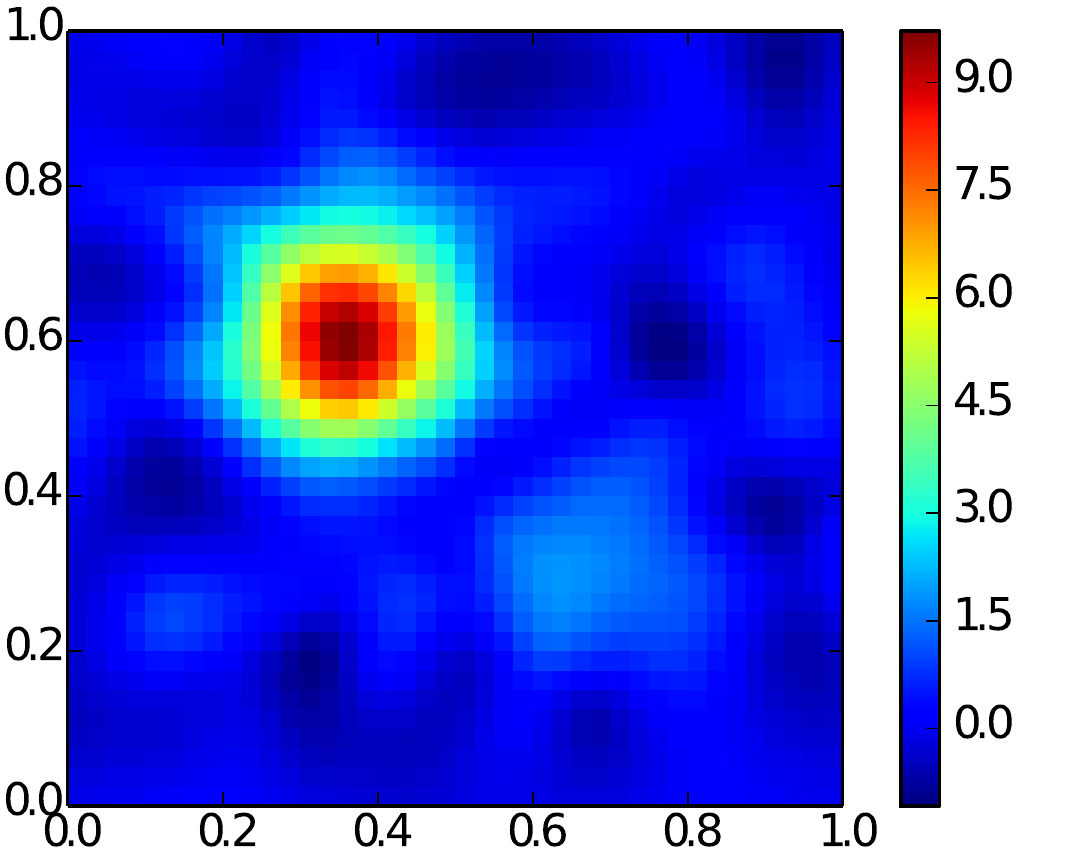}
\caption{$\langle \rho(\x)\rangle$}
\label{fig:fig2g}
\end{subfigure}
\begin{subfigure}[b]{0.32\textwidth}
\includegraphics[height=9pc]{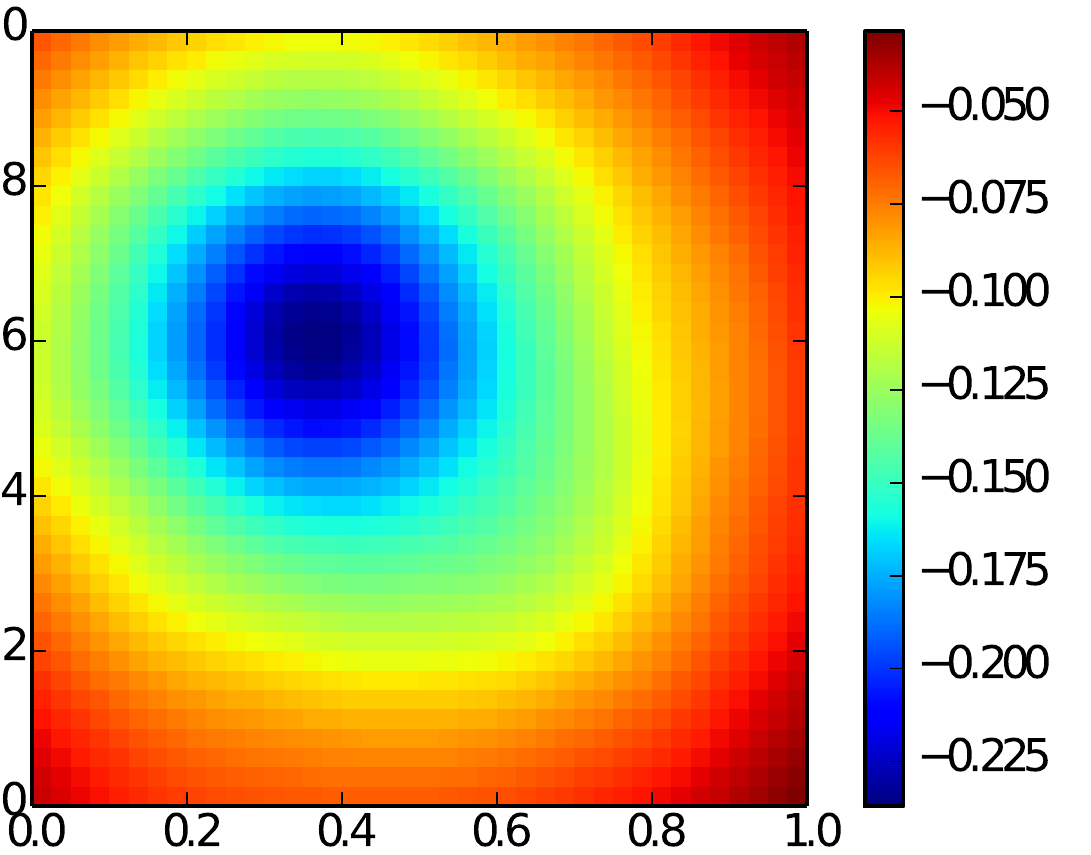}
\caption{$\langle \varphi(\x) \rangle$}
\label{fig:fig2h}
\end{subfigure}
\begin{subfigure}[b]{0.32\textwidth}
\hspace{-5pt}\includegraphics[height=9pc]{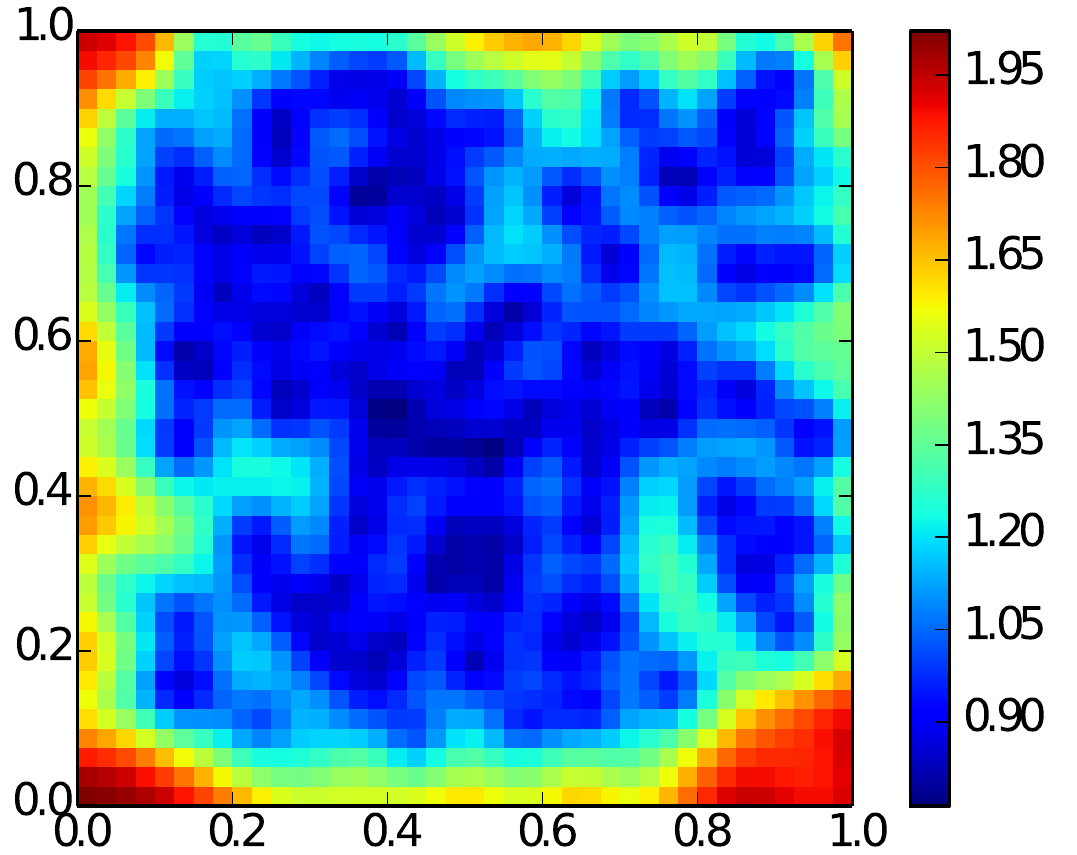}
\caption{$\sqrt{\langle (\rho(\x)-\langle \rho(\x)\rangle)^2\rangle}$}
\label{fig:fig2i}
\end{subfigure}

\caption[\textbf{Source inversion}]{\textbf{Source inversion for Poisson's equation in isotropic medium in $\mathbb{R}^2$.} 
(\subref{fig:fig2source}) Synthetic source function that is the object of recovery. (\subref{fig:fig2varphi}) 
A solution to Poisson's equation on the infinite domain corresponding
to this source. (\subref{fig:fig2obs}) $125$ randomly-placed
observations of $\varphi$ taken with noise.
(\subref{fig:fig2g}) Reconstruction of the source. (\subref{fig:fig2h}) 
Reconstruction of the potential.
(\subref{fig:fig2i}) Pointwise standard error in the 
reconstruction of the source. Parameters used: $\beta=10^{-4},
s_m=10^{-3}, \gamma = 10^2$. }
\label{fig:fig2}
\end{figure}

As an example, we recover the source function in $\mathbb{R}^2$ shown
in Fig.~\ref{fig:fig2source}. This source was used along with a
uniform unit dielectric coefficient to find the solution for the
Poisson equation that is given in Fig.~\ref{fig:fig2varphi}. Noisy samples
of the potential field were taken at $125$ randomly-placed locations (depicted in Fig.~\ref{fig:fig2obs}).
%
%
For regularization, we sought solutions for $\rho$ in the Sobolev
space $H^{2}(\mathbb{R}^2)$. Such spaces are associated with the
Bessel potential operator $P(-\Delta) = \beta(\gamma-\Delta)^2.$
%
Using $125$ randomly placed observations, reconstructions of both
$\varphi$ and $\rho$ were performed. The standard error of the
reconstruction is also given.

\subsection{Recovery of a spatially-varying dielectric coefficient field}\label{sec:dielectric}

Finally, consider the recovery of a spatially varying dielectric coefficient
$\epsilon(\x)$ by inverting the Poisson equation
\begin{equation}
\nabla\cdot\left(\epsilon\nabla \varphi\right) - \rho =0,\label{eq:poisson2}
\end{equation}
where $\rho$ is now known, and $\varphi$ is measured.  This problem is
more difficult than the problems in the previous sections. While Eq.~\ref{eq:poisson2} is bilinear
in $\epsilon$ and $\varphi$, the associated inverse problem of the recovery of $\epsilon$
given measurements of $\varphi$ is nonlinear, since $\epsilon$ does not
relate linearly to data in $\varphi$. This situation is also exacerbated by the fact
that no closed-form solution for $\epsilon$ as a function of $\varphi$ exists.

Assuming that the gradient of the dielectric coefficient is spatially
correlated according to the Gaussian process given by $P(-\Delta)$, we
work with the Hamiltonian

\begin{align}
H[\varphi,\epsilon,\lambda ; \rho, \phiobs] &= \frac{1}{2}\sum_{m=1}^M\int \frac{\delta(\x-\x_m)}{s_m^2}|\varphi(\x)-\phiobs(\x)|^2\d\x -\frac{1}{2} \int \epsilon(\x) \Delta P(-\Delta)\epsilon(\x) \d\x \nonumber\\
&\qquad\qquad+ i\int \lambda(\x)\left[ \nabla\cdot\left(\epsilon\nabla \varphi\right) - \rho \right]\d\x,\label{eq:Huepslambda}
\end{align}
which yields the Euler-Lagrange equations

\begin{align}
\nabla\cdot\left(\epsilon\nabla \varphi\right) -\rho &=0, \label{eq:poissonconstraint}\\
-\Delta P(-\Delta)\epsilon -\nabla\lambda\cdot\nabla \varphi & = 0,\label{eq:epspois}\\
\sum_{j=1}^M \frac{\delta(\x-\x_j)}{s_j^2}(\varphi(\x)-\phiobs(\x)) 
+\nabla\cdot\left(\epsilon\nabla\lambda\right) & = 0\label{eq:upois}.
\end{align}
We have assumed that $\epsilon$ is sufficiently regular such
that $\int \nabla\epsilon(\x)\cdot P(-\Delta)\nabla\epsilon(\x)\d\x <\infty$,
thereby imposing vanishing boundary-conditions at $|\x|\to\infty$. The
Lagrange multiplier $\lambda$ satisfies the Neumann boundary
conditions $\nabla \lambda=0$ outside of the convex hull of the
observed points. In order to recover the optimal $\epsilon$, one must
solve these three PDEs simultaneously. A general iterative strategy for solving this system of partial
differential equations is to use Eq.~\ref{eq:poissonconstraint} to
solve for $\varphi$, use Eq~\ref{eq:epspois} to solve for $\epsilon$,
and use Eq.~\ref{eq:upois} to solve for $\lambda$. Given $\lambda$ and
$\varphi$, the left-hand-side of Eq~\ref{eq:epspois} 
provides the gradient of the Hamiltonian with respect to $\epsilon$ which can
be used for gradient descent.
Eqs.~\ref{eq:poissonconstraint} and~\ref{eq:upois} are simply the
Poisson equation.

For quantifying error in the mean-field recovery, we seek a
formulation of the problem of recovering $\epsilon$ using the path
integral method. We are interested in the generating functional $Z[J]
= \iiint \mathcal{D}\varphi\mathcal{D}\epsilon\mathcal{D}\lambda 
\exp\left(-H[\varphi,\epsilon,\lambda]+\int
J \epsilon \d\x \right).$ Integrating in $\lambda$ and $\varphi$,
yields the marginalized generating functional

{\small
\begin{align}
\lefteqn{Z[J]=\int \mathcal{D}\epsilon\exp\left\{-H[\epsilon ; \rho, \phiobs] +\int J(\x)\epsilon(\x)\d\x \right\}}\nonumber\\
&\quad= \int \mathcal{D}\epsilon\exp\Bigg\{- \frac{1}{2}\sum_{m=1}^M\int \frac{\delta(\x-\x_m)}{s_m^2}\left[ \varphi(\epsilon(\x))-\phiobs(\x)\right]^2\d\x \nonumber\\
& \qquad-\frac{1}{2} \int \epsilon(\x)(-\Delta) P(-\Delta)\epsilon(\x) \d\x+\int J(\x)\epsilon(\x)\d\x\Bigg\}.\label{eq:Hepsimplicit1}
\end{align}}
To approximate this integral, one needs an expression for the
$\varphi$ as a function of $\epsilon$. To find such an expression, one
can use the product rule to write Poisson's equation as
$\epsilon\Delta \varphi + \nabla \epsilon\cdot\nabla \varphi =\rho.$
Assuming that $\nabla\epsilon$ is small, one may solve Poisson's
equation in expansion of powers of $\nabla\epsilon$ by using the
Green's function $L(\x,\xp)$ of the Laplacian operator to write $
\varphi(\x)=\int L(\x,\xp)\frac{\rho(\xp)}{\epsilon(\xp)}\d\xp 
-\int L(\x,\xp)\nablap\log \epsilon(\xp) \cdot 
\nablap \varphi(\xp) \d\xp $, which is a Fredholm 
integral equation of the second kind. The function $\varphi$ then has the
Liouville-Neumann series solution
\begin{align}
\varphi(\x)&= \sum_{n=0}^\infty \varphi_n(\x) \label{eq:poisson1}  \\
\varphi_n(\x)&= \int K(\x,\y) \varphi_{n-1}(\y)\d\y\qquad n\geq1\label{eq:lnpoisson} \\
\varphi_0(\x)&=\int L(\x,\y)\frac{\rho(\y)}{\epsilon(\y)}\d\y \\
K(\x,\y)&= \nabla_\y\cdot\Big[ L(\x,\y)\nabla_\y\log \epsilon(\y)\Big],\label{eq:poisson5} 
\end{align}
where $\nabla \epsilon$ is assumed to vanish at the boundary of reconstruction.  Taken
to two terms in the expansion of $\varphi(\epsilon)$ given in
Eqs.~\ref{eq:poisson1}-\ref{eq:poisson5}, the second-order term in the
Taylor expansion of Eq.~\ref{eq:Hepsimplicit1} is of the form
(see Appendix~\ref{sec:dielectricexpansion}) $$\frac{\delta^2
H}{\delta \epsilon(\x)\delta\epsilon(\xp)}\sim -\Delta
P(-\Delta)\delta(\x-\xp) + \sum_{m=1}^M a_m(\x,\xp).$$
%
This expression, evaluated at the
solution of the Euler-Lagrange equations
$\epsilon^\star, \varphi^\star$, provides an an approximation of the
original probability density from which the posterior variance
$\left\langle \epsilon(\x)-\epsilon^\star(\x), \epsilon(\xp)-\epsilon^\star(\xp) \right\rangle
= A^{-1}(\x,\xp)$ can be estimated.  To find this inverse operator, we discretize spatially and 
compute the matrix $\mathbf{A}^{-1}_{ij} = A^{-1}(\x_i,\x_j)$,
\[
\mathbf{A}^{-1} = ( \mathbf{I}+\mathbf{G}\mathbf{A}_m^{-1})^{-1}\mathbf{G},
\]
where $\mathbf{I}$ is the identity matrix, $\mathbf{G}$ is a matrix of
values $[(-\Delta)P(-\Delta)\delta(\x,\xp)]^{-1}$, $\mathbf{A}_m^{-1}
= \left[\delta\x\sum_ma_m(\x,\xp) \right]^{-1},$ and $\delta\x$ is
the volume of a lattice coordinate. 

\begin{figure}
\centering
\begin{subfigure}[b]{0.32\textwidth}
\caption{$\epsilon(\x), \epsilon^\star(\x)$}
\includegraphics[width=\textwidth]{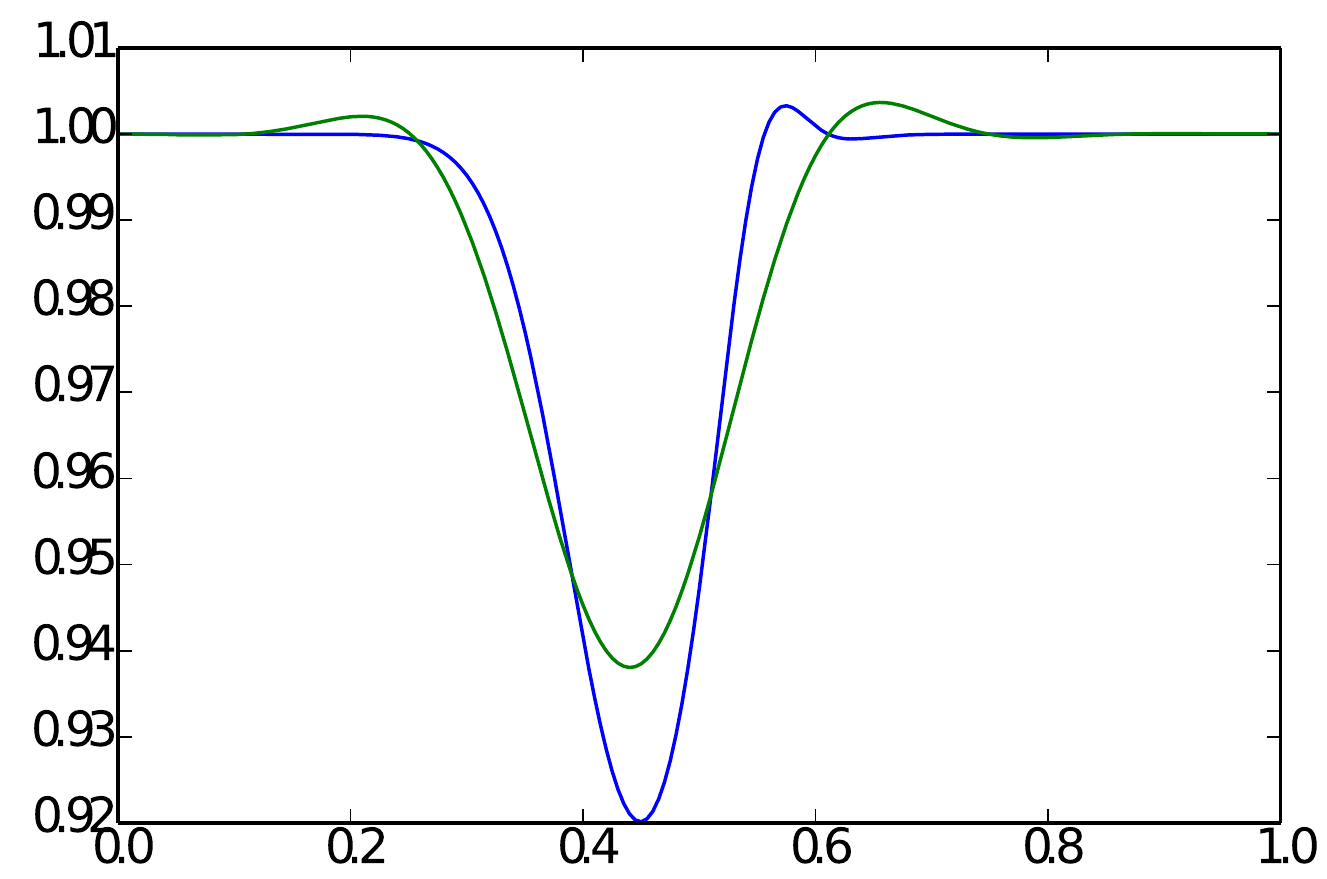}
\label{fig:epsilon}
\end{subfigure}
\begin{subfigure}[b]{0.32\textwidth}
\caption{$\varphi(\x),\phiobs(\x),\varphi^\star(\x)$}
\includegraphics[width=\textwidth]{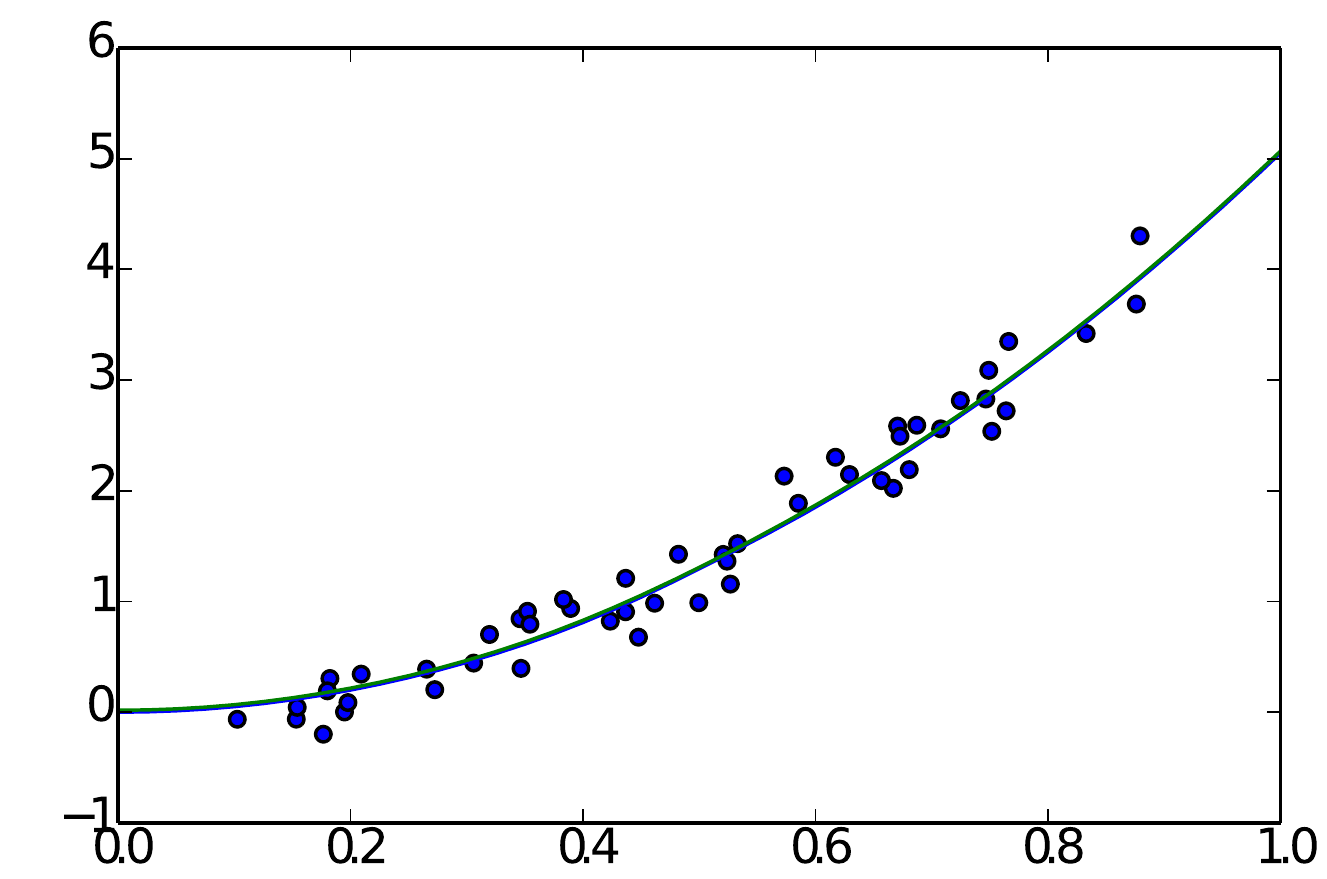}
\label{fig:varphi1d}
\end{subfigure}
\begin{subfigure}[b]{0.32\textwidth}
\caption{$\sqrt{\langle(\epsilon(\x)- \epsilon^\star(\x))^2\rangle_\textrm{approx}}$}
\includegraphics[width=\textwidth]{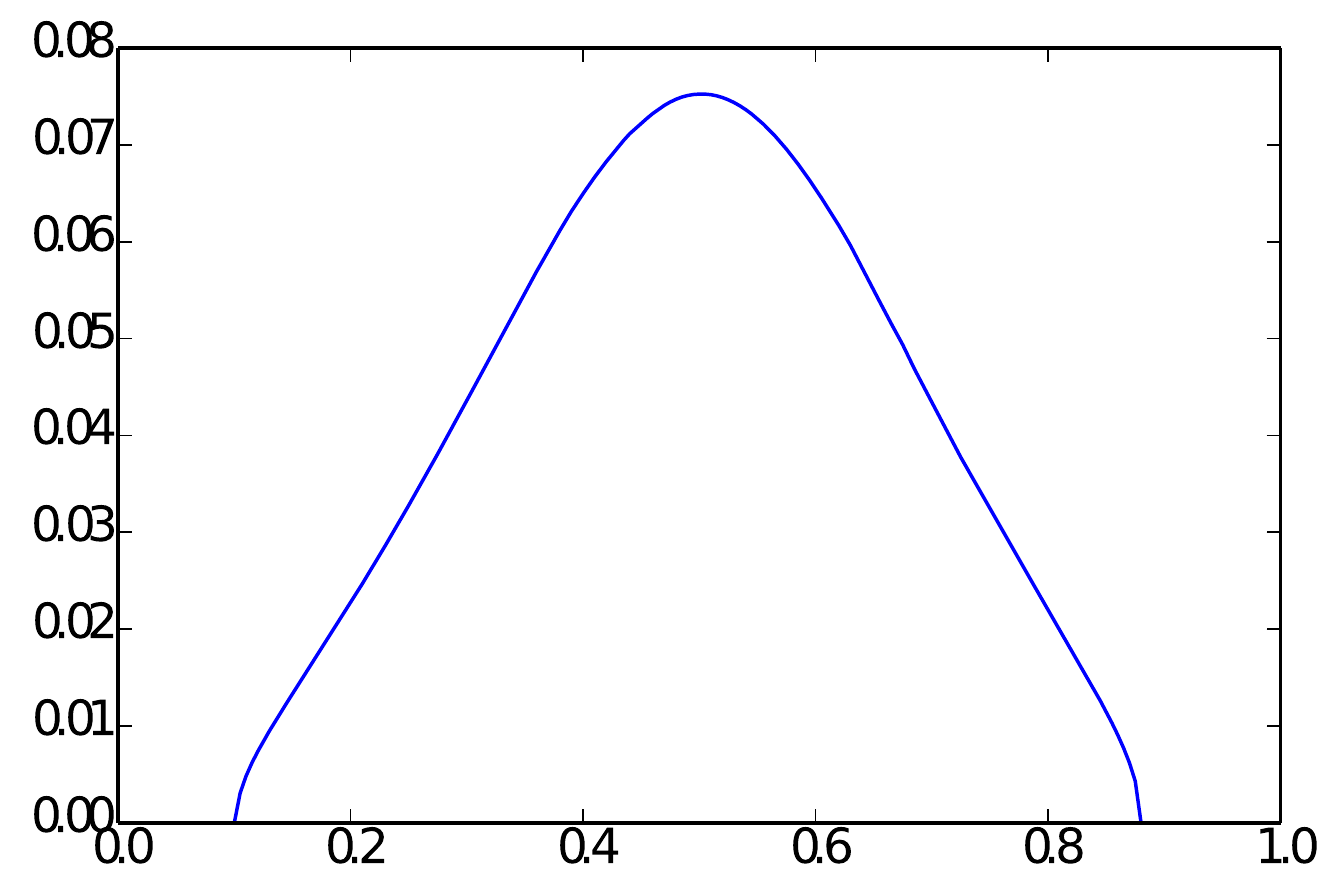}
\label{fig:epsilonerr}
\end{subfigure}
\begin{subfigure}[b]{0.32\textwidth}
\includegraphics[width=\textwidth]{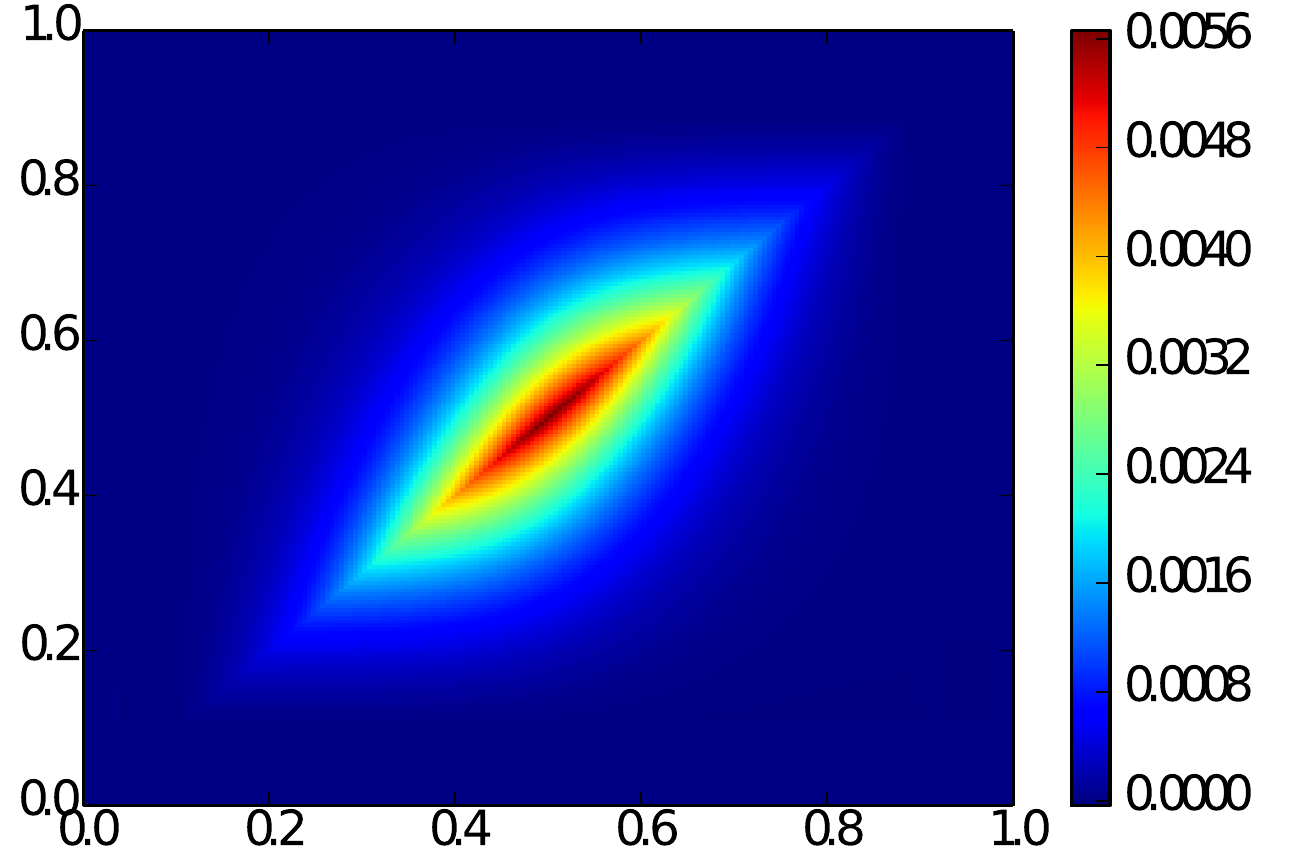}
\caption{$\langle(\delta\epsilon(\x),\delta\epsilon(\xp)\rangle_\textrm{approx} $}
\label{fig:epsilonvar}
\end{subfigure}
\begin{subfigure}[b]{0.32\textwidth}
\includegraphics[width=\textwidth]{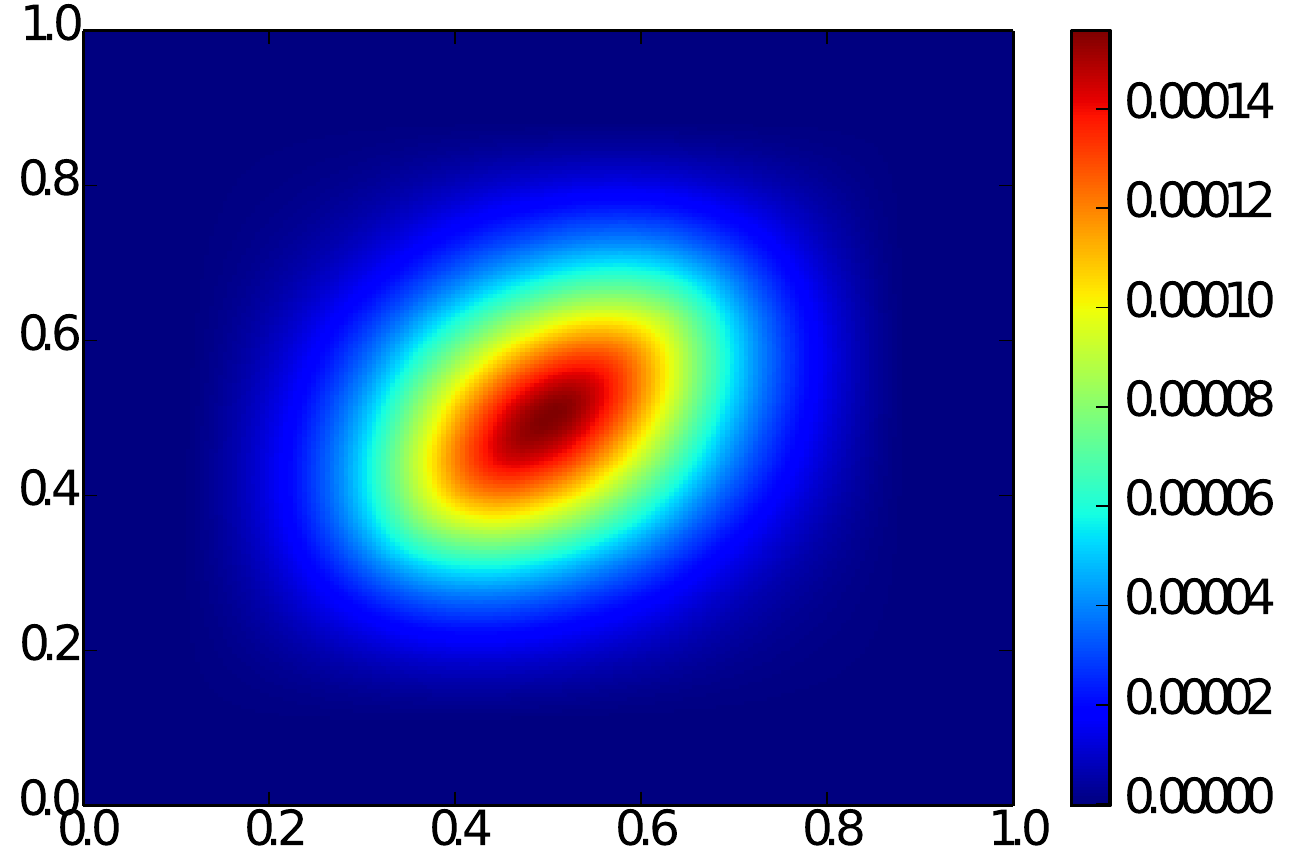}
\caption{$\langle(\delta\epsilon(\x),\delta\epsilon(\xp)\rangle_\textrm{MC}  $}
\label{fig:epsiloncovarmc}
\end{subfigure}
\begin{subfigure}[b]{0.32\textwidth}
\includegraphics[width=\textwidth]{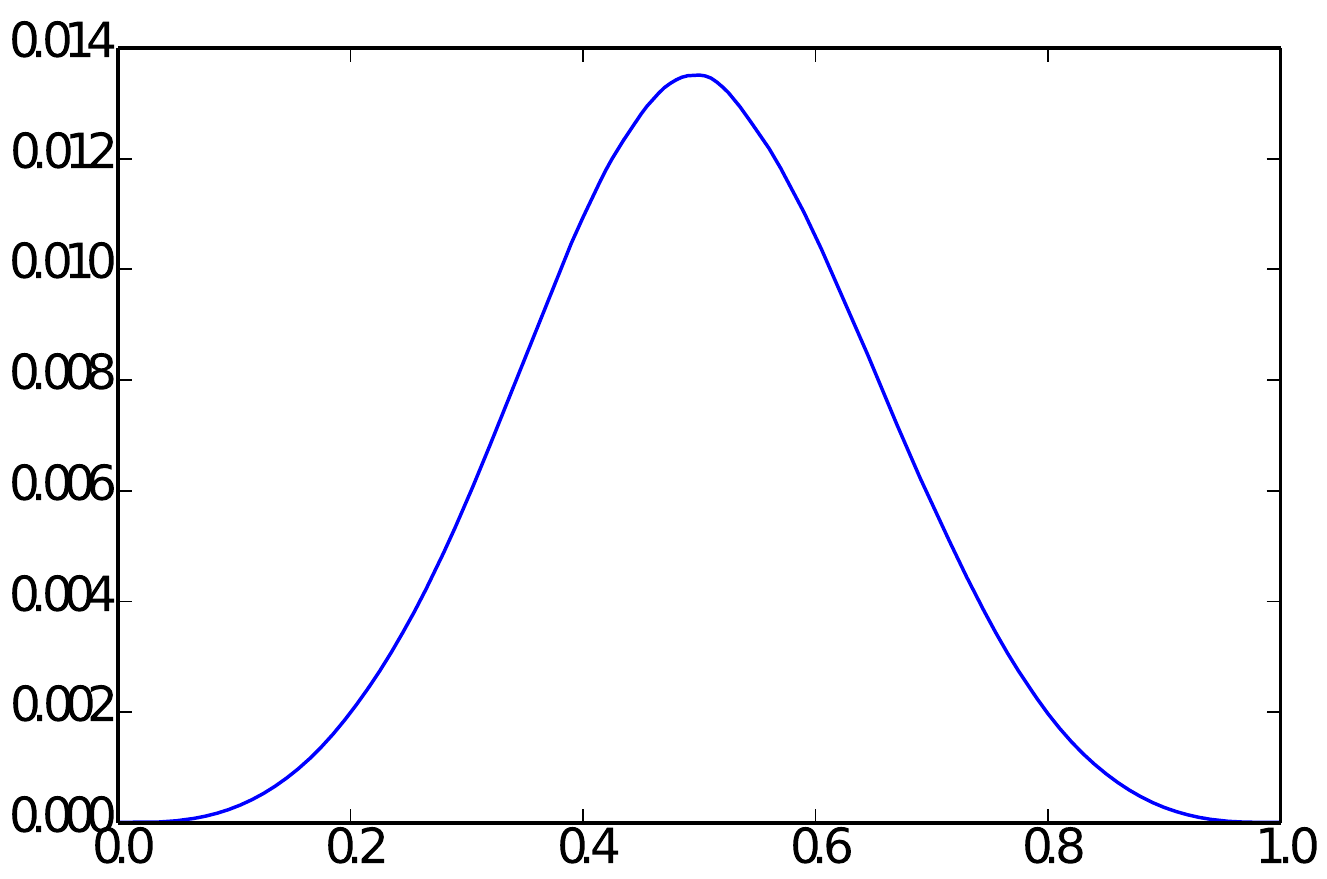}
\caption{$\sqrt{\langle(\epsilon(\x)- \epsilon^\star(\x))^2\rangle_\textrm{MC}}  $}
\label{fig:epsilonvarmc}
\end{subfigure}

\caption{\textbf{Dielectric inversion for Poisson's equation in $\mathbb{R}^1$.}  (\subref{fig:epsilon})
(blue) Spatially-varying dielectric coefficient $\epsilon$ that is the
object of recovery, and the mean field recovery $\epsilon^\star$
(green).
 (\subref{fig:varphi1d}) A solution $\varphi$ (blue) to
Poisson's equation given a known step-source supported on $[0,1]$
 and the spatially-varying dielectric coefficient, $50$ randomly placed samples of the solution taken with
error, and the mean field recovery of the potential function
$\varphi^\star$ (green).  (\subref{fig:epsilonerr}) Standard error in
the mean-field recovery of the dielectric
field. (\subref{fig:epsilonvar}) Approximate posterior variance in the
recovery of the dielectric field.  $(\delta\epsilon
= \epsilon-\epsilon^\star)$. (\subref{fig:epsiloncovarmc}) Monte-Carlo corrected covariance field estimate (\subref{fig:epsilonvarmc}) Monte-Carlo corrected point-wise error estimate. Parameters used: $s_m =
0.2, \beta=2.5, \gamma=100$.}
\end{figure}

As an example, we present the recovery of a dielectric coefficient in
$\mathbb{R}^1$ over the compact interval $x\in[0,1]$ of a dielectric
coefficient shown in Fig.~\ref{fig:epsilon} given a known
source function ($10\times\mathbf{1}_{x\in[0,1]})$. A solution to the
Poisson equation given Eq.~\ref{eq:poisson2} is shown in
Fig.~\ref{fig:varphi1d}. For regularization, we use the operator
$P(-\Delta)=\beta(\gamma-\Delta)$, and assume that $\nabla\epsilon\to0$ at the 
boundaries of the recovery, which are outside of the locations where
measurements are taken. For this reason, we take the
Green's function $G$ of the differential operator $-\frac{d^2}{dx^2}P(-\frac{d^2}{dx^2}) = -\frac{d^2}{dx^2}\beta(\gamma-\frac{d^2}{dx^2})$ to vanish along with its first two derivatives 
at the boundary of recovery.

 The point-wise standard error and
the posterior covariance are shown in Figs.~\ref{fig:epsilonerr} and 
\ref{fig:epsilonvar}, respectively. Monte-Carlo corrected estimates are also shown. Note that approximate point-wise errors are much larger than the Monte-Carlo point-wise errors. This fact
is due in-part to inaccuracy in using the series solution for the Poisson equation given
in Eq~\ref{eq:poisson1}, which relies on $\nabla\epsilon$ to be small. While the approximate errors were inaccurate, the approximation
was still useful in providing a sampling density for use in importance sampling.

\section{Discussion}
In this paper we have presented a general method for regularizing
ill-posed inverse problems based on the Bayesian interpretation of
Tikhonov regularization, which we investigated through the use of
field-theoretic approaches. We demonstrated the approach by
considering two linear problems -- interpolation
(Sec.~\ref{sec:interp}) and source inversion (Sec.~\ref{sec:source}),
and a non-linear problem -- dielectric inversion
(Sec.~\ref{sec:dielectric}). For linear problems Tikhonov
regularization yields Gaussian functional integrals, where the moments
are available in closed-form. For non-linear problems, we demonstrated a
perturbative technique based on functional Taylor series expansions,
for approximate local density estimation near the maximum a-posteriori
solution of the inverse problem. We also discussed how such
approximations can be improved based on Monte-Carlo sampling
(Sec.~\ref{sec:montecarlo}).

Our first example problem was that of membrane or plate
interpolation. In this problem the regularization term is known based
on a priori knowledge of the physics of membranes with bending
rigidity. The Helfrich free energy describes the thermal fluctuations
that are expected of rigid membranes, and provided us with the
differential operator to use for Tikhonov regularization. Using the
path integral, we were able to calculate an analytical expression for
the error in the reconstruction of the membrane surface. It is
apparent that the error in the recovery depends on both the error of
the measurements and the distance to the nearest
measurements. Surprisingly, the reconstruction error did not
explicitly depend on the misfit error.

The second example problem was the reconstruction of the source term in
the Poisson equation given measurements of the field. In this problem,
the regularization is not known from physical constraints and we
demonstrated the use of a regularizer chosen from a general family of
regularizers. This type of regularization is equivalent to the notion
of weak solutions in Sobolev spaces. Since the source inversion
problem is linear, we were able to analytically calculate the solution
as well as the error of the solution. Again, the reconstruction error
did not explicitly depend on the misfit error.

The last example problem was the inversion of the
dielectric coefficient of Poisson's equation from potential
measurements. This problem was nonlinear, yielding non-Gaussian
path-integrals. We used this problem to  demonstrate the
technique of semiclassical approximation for use in Bayesian inverse problems.

The reliability of the semiclassical approximation depends on how rapidly the posterior distribution
falls off from the extremum or mean field solution. Applying the semiclassical approximation to the Information Hamiltonian (Eq~\ref{eq:constrainedenergy}), one sees that the regularization only contributes to terms up to second order.
Higher-order terms in the expansion rely only on the likelihood term in the Hamiltonian. Since the data error is assumed to be normally distributed with variance $s_m^2$, one expects each squared residual $(\varphi(\x_m)-\varphi_\textrm{obs}(\x_m))^2$ to be $\mathcal{O}(s_m^2)$. For this reason,  each observation contributes a term of
 $O(1)$ to the Hamiltonian. As a result, there is an implicit large prefactor of $O(M)$ in 
 the Hamiltonian, where $M$ is defined as before as the number of observations. The first order correction
 to the semiclassical method is then expected to be $\mathcal{O}(1/M)$.

\subsection{Future directions}


By putting inverse problems into a Bayesian framework, one gains
access to a large toolbox of methodology that can be used to construct
and verify models. In particular, Bayesian model
comparison~\cite{o2004bayesian} methods can be used for identifying
the regularization terms to be used when one does not have prior
information available about the solution. Such methods can also be
used when one has some knowledge of the object of recovery, modulo the
knowledge of some parameters. For instance, one may seek to recover
the height of a plate or membrane but not know the surface tension or
elasticity. Then, Bayesian methods can be used to recover probability
distributions for the regularization parameters along with the object of recovery.

Finally, Tikhonov regularization works naturally in the path integral
framework because it involves quadratic penalization terms which yield
Gaussian path integrals. It would be interesting to examine other
forms of regularization over function spaces within the path integral
formulation, such as $L^1$ regularization.

\section{Acknowledgements}

This material is based upon work supported by the National Science
Foundation under Agreement No. 0635561. JC and TC also acknowledge
support from the National Science Foundation through grant
DMS-1021818, and from the Army Research Office through grant 58386MA.
VS acknowledges support from UCLA startup funds.

\appendix

\section{Functional Taylor approximations for the dielectric field problem}
\label{sec:dielectricexpansion}
We wish to expand the Hamiltonian
\begin{equation}
H[\epsilon ; \rho, \phiobs] = \frac{1}{2}\sum_{m=1}^M\int \frac{\delta(\x-\x_m)}{s_m^2}\left[\sum_{n=0}^\infty \varphi_n(\epsilon(\x))-\phiobs(\x)\right]^2\d\x +\frac{1}{2} \int \epsilon(\x)(-\Delta)P(-\Delta)\epsilon(\x) \d\x\label{eq:Hepsimplicit}
\end{equation}
about its extrema $\epsilon^\ast$. We take variations with respect to $\epsilon(\x)$ to calculate its first functional derivative,
\begin{align}
\int\frac{\partial H}{\partial \epsilon(\x)}\phi(\x)\d\x&=\int (-\Delta)P(-\Delta)\epsilon(x) \phi(\x)\d\x+\lim_{h\to0}\frac{\d}{\d h}\frac{1}{2}\sum_{m=1}^M\int \frac{\delta(\x-\x_m)}{s_m^2}\left[\sum_{n=0}^\infty \varphi_n(\epsilon(\x)+h\phi(\x))-\phiobs(\x)\right]^2\d\x  \nonumber \\
&=\int (-\Delta)P(-\Delta)\epsilon\phi \d\x+\lim_{h\to0}\sum_{m=1}^M\int\frac{\delta(\x-\x_m)}{s_m^2}\left(\varphi(\x)-\phiobs(\x) \right) \frac{\d}{\d h}\varphi_0(\epsilon(\x)+h\phi(\x))\d\x  \nonumber \\
&\qquad+\lim_{h\to0}\sum_{m=1}^M\int\frac{\delta(\x-\x_m)}{s_m^2}\left(\varphi(\x)-\phiobs(\x) \right) \frac{\d}{\d h}\varphi_1(\epsilon(\x)+h\phi(\x))\d\x  \nonumber \\
&\qquad+\underbrace{\lim_{h\to0}\sum_{m=1}^M \int \frac{\delta(\x-\x_m)}{s_m^2}\left(\varphi(\x)-\phiobs(\x) \right)\sum_{n=2}^\infty \frac{\d}{\d h} \varphi_n(\epsilon(\x)+h\phi(\x))\d\x}_{I_1} 
  \label{eq:firstvariation1}
\end{align}
Let us define the quantities
\begin{align*}
\tilde{K}(\y,\z)&=\nabla_\z\cdot\left[L(\y,\z)\nabla_\z\left(\frac{\phi(\z)}{\epsilon(\z)}\right) \right]\\
\tilde{\varphi}_0(\x) &= -\int L(\x,\y)\frac{\rho(\y)\phi(\y)}{\epsilon^2(\y)}\d\y\\
\Psi(\x)&=\sum_{m=1}^M  \frac{\delta(\x-\x_m)}{s_m^2}\left(\varphi(\x)-\phiobs(\x) \right).
\end{align*}
Through direct differentiation we find that
\begin{align*}
I_1&= \sum_{n=2}^\infty \int \Psi(\x) K(\x,\y_n) \left(\prod_{j=1}^{n-1} K(\y_{j+1}, \y_j) \right) \tilde{\varphi}_0(\y_1)\d\x\prod_{k=1}^n \d\y_k\\
&\qquad +   \sum_{n=2}^\infty \int \Psi(\x)\tilde{K}(\x,\y_n) \left(\prod_{j=1}^{n-1} K(\y_{j+1}, \y_j) \right) {\varphi}_0(\y_1)\d\x\prod_{k=1}^n \d\y_k \\
&\qquad+\sum_{n=2}^\infty \int \Psi(\x){K}(\x,\y_n) \sum_{k=0}^{n-1}\left(\tilde{K}(\y_{k+1},\y_k)\prod_{\substack{j=1\\j\neq k}}^{n-1} K(\y_{j+1}, \y_j) \right)\varphi_0(\y_1)\d\x\prod_{k=1}^n \d\y_k.
\end{align*}
Integrating in $\x$:
\begin{align*}
I_1 &=\sum_{n=2}^\infty \sum_{m=1}^M\frac{\varphi(\x_m)-\phiobs(\x_m)}{s_m^2}\int K(\x_m,\y_n) \left(\prod_{j=1}^{n-1} K(\y_{j+1}, \y_j) \right) \tilde{\varphi}_0(\y_1)\prod_{k=1}^n \d\y_k\\
&\quad +   \sum_{n=2}^\infty \sum_{m=1}^M\frac{\varphi(\x_m)-\phiobs(\x_m)}{s_m^2} \int\tilde{K}(\x_m,\y_n) \left(\prod_{j=1}^{n-1} K(\y_{j+1}, \y_j) \right) {\varphi}_0(\y_1)\prod_{k=1}^n \d\y_k \\
&\quad+\sum_{n=2}^\infty \sum_{m=1}^M\frac{\varphi(\x_m)-\phiobs(\x_m)}{s_m^2}\int K(\x_m,\y_n) \sum_{k=1}^{n-1}\left(\tilde{K}(\y_{k+1},\y_k)\prod_{\substack{j=1\\j\neq k}}^{n-1} K(\y_{j+1}, \y_j) \right)\varphi_0(\y_1)\prod_{k=1}^n \d\y_k.
\end{align*}
We shift $\phi(\cdot)\to\phi(\x)$, and integrate-by-parts to find
{\small\begin{align*}
\lefteqn{I_1 =-\sum_{n=2}^\infty \sum_{m=1}^M\frac{\varphi(\x_m)-\phiobs(\x_m)}{s_m^2}\int K(\x_m,\y_n) \left(\prod_{j=1}^{n-1} K(\y_{j+1}, \y_j) \right)L(\y_1,\x)\frac{\rho(\x)}{\epsilon^2(\x)}\phi(\x)\d\x\prod_{k=1}^n \d\y_k}\\
&\quad +   \sum_{n=2}^\infty \sum_{m=1}^M\frac{\varphi(\x_m)-\phiobs(\x_m)}{s_m^2} \int \frac{\phi(\x)}{\epsilon(\x)}\nabla\cdot\left[L(\x_m,\x)\nabla\left(K(\x,\y_{n-1})\right) \right] \left(\prod_{j=1}^{n-2} K(\y_{j+1}, \y_j) \right) {\varphi}_0(\y_1)\d\x\prod_{k=1}^n \d\y_k \\
&\quad+\sum_{n=2}^\infty \sum_{m=1}^M\frac{\varphi(\x_m)-\phiobs(\x_m)}{s_m^2}\int K(\x_m,\y_n) \sum_{k=1}^{n-1}\left(\frac{\phi(\x)}{\epsilon(\x)}\nabla\cdot\left[L(\y_{k+1},\x)\nabla K(\x,\y_{k-1}) \right]\prod_{\substack{j=1\\j\neq k}}^{n-2} K(\y_{j+1}, \y_j) \right)\varphi_0(\y_1)\d\x\prod_{k=1}^n  \d\y_k.
\end{align*}}
Note that all boundary terms disappear since we can take $\phi$ to disappear on the boundary. With $I_1$ computed, we find{\small
\begin{align}
\frac{\delta H}{\delta\epsilon(\x)}&=(-\Delta)P(-\Delta)\epsilon(\x)-\sum_{m=1}^M \frac{\varphi(\x_m)-\phiobs(\x_m)}{s_m^2}\left[ L(\x_m,\x)\frac{\rho(\x)}{\epsilon^2(\x)}   \right]\nonumber\\
&\quad + \sum_{m=1}^M \frac{\varphi(\x_m)-\phiobs(\x_m)}{s_m^2\epsilon(\x)}\nabla\cdot\left[ L(\x_m,\x)\nabla\varphi_0(\x) \right] -\sum_{m=1}^M \frac{\varphi(\x_m)-\phiobs(\x_m)}{s_m^2} \left(\frac{\rho(\x)}{\epsilon^2(\x)} \right)\int K(\x_m,\y_1)L(\x,\y_1) \d\y_1 \nonumber\\
&\quad-\sum_{n=2}^\infty \sum_{m=1}^M\frac{\varphi(\x_m)-\phiobs(\x_m)}{s_m^2}\int K(\x_m,\y_n) \left(\prod_{j=1}^{n-1} K(\y_{j+1}, \y_j) \right)L(\y_1,\x)\frac{\rho(\x)}{\epsilon^2(\x)}\prod_{k=1}^n \d\y_k \nonumber\\
&\quad +   \sum_{n=2}^\infty \sum_{m=1}^M\frac{\varphi(\x_m)-\phiobs(\x_m)}{s_m^2\epsilon(\x)} \int \nabla\cdot\left[L(\x_m,\x)\nabla\left(K(\x,\y_{n-1})\right) \right] \left(\prod_{j=1}^{n-2} K(\y_{j+1}, \y_j) \right) {\varphi}_0(\y_1)\prod_{k=1}^n\d\y_k\nonumber \\
&\quad+\sum_{n=2}^\infty \sum_{m=1}^M\frac{\varphi(\x_m)-\phiobs(\x_m)}{s_m^2\epsilon(\x)}\int K(\x_m,\y_n) \sum_{k=1}^{n-1}\left(\nabla\cdot\left[L(\y_{k+1},\x)\nabla K(\x,\y_{k-1}) \right]\prod_{\substack{j=1\\j\neq k}}^{n-2} K(\y_{j+1}, \y_j) \right)\varphi_0(\y_1)\prod_{k=1}^n  \d\y_k.
\end{align}}
Taken to two terms in the series expansion for $\varphi$, the first variation is
{\small\begin{align}
\frac{\delta H}{\delta\epsilon(\x)}&\sim (-\Delta)P(-\Delta)\epsilon(\x)+\sum_{m=1}^M \frac{\varphi(\x_m)-\phiobs(\x_m)}{s_m^2\epsilon(\x)}\left[ \nabla L(\x,\x_m)\cdot\nabla\varphi_0(\x) -\frac{\rho(\x)}{\epsilon(\x)}\int K(\x_m,\y_1)L(\x,\y_1)\d\y_1 \right].
\end{align}}

To calculate the second-order term in the Taylor-expansion, we take another variation. Truncated at two terms in the expansion for $\varphi$:
\begin{equation}
\frac{\delta^2 H}{\delta\epsilon(\x)\delta\epsilon(\xp)}= (-\Delta)P(-\Delta)\delta(\x-\xp)+\sum_{m=1}^M a_m(\x,\xp),
\end{equation}
where after canceling like terms,
{\small\begin{align*}
a_m(\x,\xp)&=\delta(\x-\xp) \frac{\varphi(\x_m)-\phiobs(\x_m)}{s_m^2\epsilon^2(\xp)}\left[ \frac{2\rho(\xp)}{\epsilon(\xp)}\int K(\x_m,\y_1)L(\xp,\y_1)\d\y_1 -\nablap L(\xp,\x_m)\cdot\nablap\varphi_0(\xp)-L(\x_m,\xp)\frac{\rho(\xp)}{\epsilon(\xp)}   \right]\nonumber \\
&\quad-\frac{\varphi(\x_m)-\phiobs(\x_m)}{s_m^2}\left\{ \nabla L(\x,\x_m)\cdot\nablap L(\x,\xp)  \frac{\rho(\xp)}{\epsilon(\x)\epsilon^2(\xp)}+\frac{\rho(\x)}{\epsilon^2(\x)\epsilon(\xp)} \nabla L(\x,\xp)\cdot\nablap L(\x_m,\xp) \right\} \\
&\quad+ \left[ \nabla L(\x,\x_m)\cdot\nabla\varphi_0(\x) -\frac{\rho(\x)}{\epsilon(\x)}\int K(\x_m,\y_1)L(\x,\y_1)\d\y_1 \right]\\
&\qquad\qquad\times\frac{1}{s_m^2 \epsilon(\x)\epsilon(\xp)}\left[ \nablap L(\xp,\x_m)\cdot\nablap\varphi_0(\xp) -\frac{\rho(\xp)}{\epsilon(\xp)}\int K(\x_m,\y_1)L(\xp,\y_1)\d\y_1  \right].
\end{align*}}
It is using this expression that we can construct an approximate probability density for our field $\epsilon$.

\bibliography{inverse}

\end{document}